%% file: Rev_dc_main.tex
\documentclass[footinbib,twocolumn,prapplied]{revtex4-2}

\usepackage{graphicx}% Include figure files
\usepackage{bm}% bold math
\usepackage{color}
\usepackage{soul}
\usepackage{amsmath,amsfonts,amssymb}
\usepackage{physics}
\usepackage{extarrows} 
\usepackage[scr=boondoxo,scrscaled=1.05]{mathalfa}
\DeclareMathAlphabet\mathrsfso{U}{rsfso}{m}{n}
\newcommand{\nc}{\newcommand}
\nc{\nn}{\nonumber}
\def\e{\mathcal{E}}
\def\U{\mathcal{U}}

\nc{\XYZ }{}
\usepackage[normalem]{ulem}

\nc{\rev }{\color{red} }

\begin{document}
\title{Electronic-based model of the optical nonlinearity of low-electron-density-Drude materials}

\date{\today}
\author{Ieng-Wai Un}
\email{iengwai@post.bgu.ac.il}
\thanks{These two authors have contributed equally.}
\author{Subhajit Sarkar}
\email{subhajit@post.bgu.ac.il}
\thanks{These two authors have contributed equally.}
\author{Yonatan Sivan}
\email{sivanyon@bgu.ac.il}
\affiliation{School of Electrical and Computer Engineering, Ben-Gurion University of the Negev and the Ilse Katz Center for Nanoscale Science and Technology, Ben-Gurion University of the Negev, Beer Sheva, Israel}

\begin{abstract}
Low electron density Drude (LEDD) materials such as indium tin oxide (ITO) are receiving considerable attention for their combination of CMOS compatibility, unique epsilon-near-zero (ENZ) behavior, and giant ultrafast nonlinear thermo-optic response.
%The understanding of low electron density Drude (LEDD) materials, such as indium tin oxide (ITO), has been limited due to modeling that only extended known models of noble metals without considering the interplay of factors such as lower electron density, high Debye energy, and non-parabolic band structure
{\XYZ However, current understanding of the electronic and optical response of LEDD materials is limited due to the simplistic modeling that only extends the known models of noble metals 
%frequently 
without considering the interplay among the lower electron density, relatively high Debye energy, and the non-parabolic band structure. We bridge this knowledge gap and provide} 
%To bridge this knowledge gap, this work provides
a complete understanding of the nonlinear electronic-thermal-optical response of LEDD materials. In particular, we rely on state-of-the-art electron dynamics modeling, as well as a time-dependent permittivity model for LEDD materials under optical pumping within the adiabatic approximation. We find the electron temperature may reach values much higher than realized before, even exceeding the Fermi temperature, in which case the effective chemical potential dramatically decreases and even becomes negative, thus, transient giving the material some characteristics of a semiconductor. We further show that the nonlinear optical response of LEDD materials originating from the changes to the real part of the permittivity is associated with changes of the population. This resolves the argument about the rise time of the permittivity, showing that it is instantaneous. In this vein, we show that referring to the LEDD permittivity as having a Kerr or ``saturable'' nonlinearity is unsuitable since its permittivity dynamics is absorptive rather than non-resonant and does not originate from population inversion. Finally, we analyze the probe pulse dynamics and unlike previous work, we obtain a quantitative agreement with the results of recent experiments.
\end{abstract}

%contrasting previous work that referred to the LEDD permittivity as having a Kerr or "saturable" nonlinearity.

\maketitle

\section{Introduction}
The technologically-important transparent conducting oxide ITO has recently been shown to possess an exceptionally-strong ultrafast response to illumination, making it a promising candidate for nonlinear optics applications. Its strong response was initially associated with the unique epsilon-near-zero (ENZ) point it has in the near infrared~\cite{Khurgin_ENZs_nlty} (see also Fig.~\ref{fig:scheme}), which is characteristic of Low Electron Density Drude (LEDD) materials. 

The nonlinear sub-picosecond optical response of LEDD materials was described using a variety of approaches, starting from a phenomenological temporally delayed response or as having a time-dependent effective mass (and hence, plasma frequency), see e.g.,~\cite{Guo_ITO_AM_2017,Guo_ITO_nanorod_natphoton,Boyd_Nat_Phot_2018,Kinsey_ENZ_OptMatExp,Xian_group_ITO_2020,Exeter_Nat_Comm_2021,Exeter_Nat_Comm_2021,Sapienza_2022,de_Leon_Mexicans_2022_1}. The latter is a simple description of the transient occupation of high energy electron states for which the effective mass is higher than for low energy states. 

More recent models employed a Relaxation Time Approximation (RTA)-based extended Two Temperature Model (eTTM)~\cite{Boyd_NLO_ENZ_ITO,Boyd_Nat_Phot_2018,de_Leon_Mexicans_2022_1,Ellenbogen-Minerbi-ITO}. Notably, while such models offered qualitative agreement with measurements, reaching also a quantitative match required various ad hoc corrections see e.g.,~\cite{de_Leon_Mexicans_2022_1}. This could have originated from the use of equations derived for parabolic bands, because the importance of momentum conservation to the $e$-$ph$ interactions was not yet understood~\cite{Un-Sarkar-Sivan-LEDD-I}, because the $e$-$e$ interaction were accounted for in a simplistic manner~\footnote{In particular, the dependence of the $e$-$e$ collision rate on the energy difference of the colliding electrons was fixed to the absorbed photon frequency, thus, overestimating the thermalization rate; the $e$-$e$ collision rate is anyhow quite fast, this is likely to have a small effect on the results.}, because the effect of the phonon temperature on the ITO permittivity was not always accounted for, or because the analysis of the nonlinear response was specific for every experimental configuration. Thus, for example, the computed eTTM parameters in~\cite{de_Leon_Mexicans_2022_1} (including the strength of the $e$-$ph$ interaction and the $T_e$-dependence of the damping factor) had to be manually changed to fit the experimental data, without a firm theoretical justification.

Notably, there are two additional limitations of the state-of-the-art nonlinear optical theory of LEDD materials. First, an important aspect of the dynamics that is not yet well understood is the response to high illumination intensities. In~\cite{Boyd_Nat_Phot_2018,Xian_group_ITO_2019,Sapienza_2022}, this response was referred to as saturable; the saturation intensity was claimed to be $\sim 100$~GW/cm$^2$~\cite{Sapienza_2022}, but this value was not connected to a population inversion (as in atomic media) nor to a balance of excitation and relaxation mechanisms, as done for Drude metals~\cite{Dubi-Sivan-Faraday}. Indeed, usually, a nonlinear saturable response is associated with the gradual depletion of the (electronic) ground state of the material, an effect which leads to a {\em decrease} of the imaginary part of its permittivity (saturable absorption)~\cite{Boyd-book}, and even to a change of its sign (upon population inversion)~\cite{Ziolkowsky_nanoshell,Sivan-OE-gain-2009}. This behavior is in contradiction to the experimental results that show that the imaginary part of the ITO permittivity {\em grows} upon illumination, see e.g.,~\cite{Xian_group_ITO_2019,de_Leon_Mexicans_2022_1}. Thus, to understand whether the response of LEDD materials is saturable or not requires simultaneous monitoring of the population as well as of the imaginary part of the permittivity via electronic simulations. 

Second, the strong changes of the permittivity necessarily lead to changes of the local field, which then affects the distribution and permittivity and vice versa. To date, the coupling between these properties was not treated self-consistently. 

A step towards resolution of the above questions was made in~\cite{Un-Sarkar-Sivan-LEDD-I} where the electron distribution dynamics was studied by solving the Boltzmann equation and extracting from it the (effective) electron and phonon temperatures, as well as the underlying thermal properties such as the heat capacity and $e$-$ph$ coupling. This work revealed the importance of momentum conservation in $e$-$ph$ interactions, the faster $e$-$e$ collision rates, and large quantitative differences in the values of various parameters compared to noble metals. Notably, however, the simulations in~\cite{Un-Sarkar-Sivan-LEDD-I} were limited to modestly high intensities.

In this work, we go beyond~\cite{Un-Sarkar-Sivan-LEDD-I} and provide a detailed electronic model of the optical properties of LEDD materials and apply it to illumination levels reaching the reported damage threshold. In Section~\ref{sec:formulation}, we recall briefly the electronic model developed in~\cite{Un-Sarkar-Sivan-LEDD-I}, and describe the time-dependent permittivity model used in this work. In particular, we model the permittivity by applying the Lindhard formula to the nonparabolic band characteristic of LEDD materials %using an extension of the Lindhard formula, 
and describe a self-consistent approach for the calculation of the distribution, permittivity and local field. %To our knowledge, this provides the first ever extension of the Lindhard formula to materials with a non-parabolic conduction band under high-intensity pulsed illumination.

As a specific example, we then focus on a prototypical geometry of an ITO layer illuminated by an obliquely incident pump pulse. In Section~\ref{sec:T-dynamics}, we discuss the resulting dynamics of the electron and phonon temperatures as a function of illumination intensity. We show that the electron temperature may reach values much higher than realized before, even exceeding the Fermi temperature, and the chemical potential may become negative, giving the material some transient characteristics of a semiconductor. We also show that the decay rate of the electron temperature becomes faster with the illumination intensity. 

In Section~\ref{subsec:epsilon-dynamics}, we study the permittivity dynamics in detail, including the various contributions to the change of its real and imaginary parts. We show that the dynamics is dominated by the change of the real part of the permittivity (causing a frequency shift of the ENZ point) %spectrally shifting the resonance away from the ENZ point)
which itself is dominated by changes to the electron distribution. The increase of the imaginary part of the permittivity is also large, but not nearly as much as that of the real part; its dynamics results from (sometimes opposite) contributions of various effects and both electron and phonon temperatures. The former insight explains the observation of the slower decay rate of the reflectivity with intensity reported in~\cite{Sapienza_2022}. Most importantly, these results show that the ITO permittivity dynamics has an unusual absorptive yet instantaneous response which is responsible for the turn on stage of the nonlinearity, whereas the turn off stage has a thermal nature, yet a relatively fast one (in agreement with~\cite{Khurgin_Kinsey_LPR}). Our results also show that the response has some characteristics of a high quality metal (as discussed in~\cite{Stoll_review} for pulsed illumination, and even in~\cite{Gurwich-Sivan-CW-nlty-metal_NP,IWU-Sivan-CW-nlty-metal_NP} for CW illumination), so that although the local field grows sublinearly with illumination intensity, the ITO is {\em not} a saturable absorber (for which the real part of the permittivity hardly changes and its imaginary part decreases with growing illumination intensity, but rather exhibits the opposite behaviour). By comparing our simulation results with the Two Temperature Model (TTM), we show that the electron dynamics described by the TTM is an excellent approximation for calculating the nonlinear optical response of ITO when the pulse duration is longer than the relatively short thermalization time. We also explain the physical origin of the ad hoc corrections needed previously, e.g., in~\cite{de_Leon_Mexicans_2022_1}. Then, in Section~\ref{subsec:eng_partition}, we discuss the energy partition and show that our prediction for the intensity at which the phonon temperature reaches the melting temperature matches well the experimentally reported damage threshold. In Section~\ref{subsec:probe_dynamics} we study the probe pulse dynamics. In particular, we show that the decay rate of the reflection of the probe pulse decreases with the pump pulse intensity due to the frequency shift of the ENZ point. This result provides quantitative agreement with the experimental observation in~\cite{Sapienza_2022}. We also study the nonlinear optical response to shorter pulses in Section~\ref{subsec:short_pulse} and show that one needs to go beyond the TTM only when the pulse duration is shorter than the thermalization time, by accounting for the non-thermal part of the distribution. Finally, we provide a discussion of the results in Section~\ref{sec:discussion} and the conclusion in Section~\ref{sec:conclusions}. 

\section{Formulation}\label{sec:formulation}
\subsection{Model for the electron distribution dynamics and phonon temperature}
We start by revisiting and extending the theoretical approach presented in~\cite{Un-Sarkar-Sivan-LEDD-I}. In particular, we solve the Boltzmann equation (BE) for the electron dynamics, namely,
\begin{widetext}
\begin{multline}\label{eq:f_neq_dynamics}
\dfrac{\partial f(\e,t)}{\partial t} = \left(\dfrac{\partial f(\e,t;\varepsilon(t,\omega_\text{pump}),|{\bf E}(t,\omega_\text{pump})|^2)}{\partial t}\right)_{\text{exc}} + \left(\dfrac{\partial f(\e,t)}{\partial t}\right)_{e\text{-}ph\ \text{collision}} \\ + \left(\dfrac{\partial f(\e,t)}{\partial t}\right)_{e\text{-}e\ \text{collision}} + \left(\dfrac{\partial f(\e,t)}{\partial t}\right)_{e\text{-}imp\ \text{collision}}.
\end{multline}
\end{widetext}
Here, $f(\e,t)$ is the electron distribution function at an energy $\e$ and time $t$, representing the population probability of electrons in a system characterized by a continuum of electron energy states within the conduction band. While the expressions for the electron-electron ($e$-$e$), electron-phonon ($e$-$ph$) and electron-charged-impurity ($e$-$imp$) interactions are the same as in~\cite{delFatti_nonequilib_2000,Seidman-Nitzan-non-thermal-population-model,Dubi-Sivan,Un-Sarkar-Sivan-LEDD-I}, the excitation term is more complicated. In particular, the electron population evolves in time such that it is given by
\begin{widetext}
\begin{multline}\label{eq:dfEdt_exc}
\left(\dfrac{\partial f(\e,t))}{\partial t}\right)_\text{exc} = B(t;\omega_\text{pump}) \Big[D_J(\e - \hbar\omega_\text{pump},\e) \rho_e(\e - \hbar\omega_\text{pump}) f(\e - \hbar\omega_\text{pump},t))(1 - f(\e,t)) \\
- D_J(\e,\e + \hbar\omega_\text{pump}) \rho_e(\e + \hbar\omega_\text{pump}) f(\e,t))(1 - f(\e + \hbar\omega_\text{pump},t))\Big],
\end{multline}
\end{widetext}
where $D_J(\e_\text{initial},\e_\text{final})$ is the squared magnitude of the transition matrix element for the electronic process $\e_\text{initial} \rightarrow \e_\text{final}$. The constant $B(t;\omega_\text{pump})$ is determined by ensuring that the increase rate of the energy density of the electron subsystem due to the excitation is equal to the absorbed power density $p_\text{abs}$, namely,
\begin{align}\label{eq:exc_eng_conserv}
\int \e \rho_e(\e) \left(\dfrac{\partial f(\e,t)}{\partial t}\right)_\text{exc} d\e = p_\text{abs}(t;\omega_\text{pump}). 
\end{align}
Here, the absorbed power density is evaluated dynamically and {\XYZ self-consistently with $f(\mathcal{E},t)$, $\varepsilon(t;\omega_\text{pump})$ and ${\bf E}(t)$} using Poynting's theorem (see details in Section~\ref{sec:scf_epsilon}). 

The BE~(\ref{eq:dfEdt_exc}) is complemented by a coarse-grained equation for the phonon temperature, which follows from energy exchange between the electron and phonon subsystems (as done previously in~\cite{Dubi-Sivan,Sarkar-Un-Sivan-Dubi-NESS-SC}), see Appendix~\ref{app:eTTMnTTM}; energy transfer to the environment occurs on the time scale of many picoseconds, thus, can be ignored in the context of the ultrafast dynamics.

These coupled equations are now used to determine the permittivity dynamics. 

\subsection{Self-consistent field calculations and model for a time-dependent permittivity}
\label{sec:scf_epsilon}
When the pump pulse energy is high enough to induce a non-negligible dynamical change in the electron distribution, the electron system cannot be considered to be time translation invariant. In this case, one cannot apply the convolution theorem to the relation between the polarization vector and the electric field vector. Strictly speaking, this requires one to solve Maxwell’s equations coupled with the Boltzmann equation using a time step much finer than the periodicity of the electromagnetic wave. Such calculation can be very time consuming and computationally expensive~\footnote{For the example analyzed in this work where the carrier frequency is 230 THz, a time-resolution of $< 0.2$ fs would be required. }. However, as shown in Appendix~\ref{app:adiab_approx}, this problem can be circumvented by applying the adiabatic approximation to calculate the time evolution of the electric field when the instantaneous intensity of the incoming pulse is weak enough such that the change rate of the electron distribution is slower than the damping rate. This condition is satisfied when the local field is smaller than $\sim 2.7\times 10^9$ V/m. For this purpose, we further assume that the pump pulse duration is much longer than the periodicity so that the electric field $\bm{\mathscr{E}}(t)$ (and the electric displacement $\bm{\mathscr{D}}(t)$) can be considered quasi-monochromatic and be written as a product of a slowly varying envelope (${\bf E}(t)$ or ${\bf D}(t)$) and a rapidly varying phase factor, namely, $\bm{\mathscr{E}}(t) = {\bf E}(t) e^{-i\omega_\text{pump}t} + \text{c.c.}$ ($\bm{\mathscr{D}}(t) = {\bf D}(t) e^{-i\omega_\text{pump}t} + \text{c.c.}$). We show in Appendix~\ref{app:adiab_approx} that the electric displacement envelope is then related to the local field envelope by ${\bf D}(t) = \varepsilon_0 \varepsilon(t,\omega_\text{pump}) {\bf E} (t)$, and that Amp\`{e}re's law and Maxwell–Faraday equation become
\begin{align}\label{eq:Maxwell_eq_adiab}
\begin{cases}
\nabla\times{\bf H}({\bf r},t) = - i \omega_\text{pump} \varepsilon_0 \varepsilon(t;\omega_\text{pump}) {\bf E}({\bf r},t) \\
\nabla\times{\bf E}({\bf r},t) = - i \omega_\text{pump} \mu_0 {\bf H}({\bf r},t)
\end{cases}.
\end{align}
Here, $\varepsilon(t;\omega_\text{pump})$ is the time-dependent permittivity given by
\begin{widetext}
\begin{align}\label{eq:eps_pump_adiab}
\varepsilon(t;\omega_\text{pump}) = \varepsilon_\infty + 
\lim_{{\bf q} \rightarrow 0} \dfrac{2e^2}{\varepsilon_0 q^2} \int \dfrac{d^3k}{(2\pi)^3} \dfrac{f_{{\bf k} + {\bf q}}(t;\omega_\text{pump}) - f_{{\bf k}}(t;\omega_\text{pump})}{\e_{{\bf k} + {\bf q}} - \e_{{\bf k}} - \hbar\omega_\text{pump} - i\hbar(\eta_{{\bf k} + {\bf q}}(t;\omega_\text{pump}) + \eta_{{\bf k}}(t;\omega_\text{pump}))/2},
\end{align}
\end{widetext}
where $\varepsilon_\infty$ represents the contribution of interband transitions to the permittivity, $f_{\bf k}$ is the momentum (${\bf k}$)-dependent electron distribution in the conduction band, $\e_{{\bf k}}$ is the electron energy satisfying the $\e$-$k$ relation (see Eq.~\eqref{eq:e-k} below), ${\bf q}$ is the wavevector of the applied field (taken to zero limit because it is much smaller than the electron wavevector), and $\eta_{{\bf k}} = \tau^{-1}_{e\text{-}e,{\bf k}} + \tau^{-1}_{e\text{-}ph,{\bf k}} + \tau^{-1}_{e\text{-}imp,{\bf k}}$ is the damping rate following from Matthiessen's rule~\cite{Ashcroft-Mermin}. $\tau_{e\text{-}e,{\bf k}}^{-1}$, $\tau_{e\text{-}ph,{\bf k}}^{-1}$ and $\tau_{e\text{-}imp,{\bf k}}^{-1}$ are the collision rates associated with the $e$-$e$, $e$-$ph$ and $e$-$imp$ interactions, respectively (see details in~\cite{Un-Sarkar-Sivan-LEDD-I}). In addition, the contribution of the interband transition ($\varepsilon_\infty$) is assumed to be independent of time, temperature and electric field and not to contribute to the dispersion at frequencies below the interband threshold of $\sim 3.2$ eV.

The local field envelope (the solution of Eqs.~\eqref{eq:Maxwell_eq_adiab}) and the time-dependent permittivity~\eqref{eq:eps_pump_adiab} are next used to calculate the power absorbed density via Poynting's theorem
\begin{align}\label{eq:P_abs}
p_\text{abs}(t;\omega_\text{pump}) = \dfrac{\omega_\text{pump}}{2} \varepsilon_0 \varepsilon''(t;\omega_\text{pump})|{\bf E}(t;\omega_\text{pump})|^2.
\end{align}
Then, we substitute the absorbed power density back to Eqs.~\eqref{eq:dfEdt_exc} and~\eqref{eq:exc_eng_conserv} to evaluate the electron-photon excitation term so that the electron dynamics and the pump pulse dynamics are solved self-consistently. Consequently, we can define Eqs.~\eqref{eq:f_neq_dynamics}-\eqref{eq:P_abs} as an Adiabatic Non-Thermal permittivity ($\mathrsfso{E}$) Model (ANTH$\mathrsfso{E}$M).  

It is then customary to rewrite the integral in energy space. For materials with a non-parabolic conduction band, e.g., in ITO, the energy-momentum ($\e$-$k$) relation is expressed by the Kane quasi-linear dispersion~\cite{Kane-quasilinear}, 
\begin{align}\label{eq:e-k}
\hbar^2 k^2 = 2 m_e^\ast \e_{{\bf k}}(1 + C\e_{{\bf k}}),
\end{align}
such that the electron density of states (eDOS) is given by~\cite{Liu_ITO_2014,Kane-quasilinear,Un-Sarkar-Sivan-LEDD-I}
\begin{align}\label{eq:DOS}
\rho_e(\e) = \dfrac{1 + 2C\e}{2\pi^2}\left(\dfrac{2 m_e^\ast}{\hbar^2}\right)^{3/2} \sqrt{\e(1 + C\e)},
\end{align}
where $m_e^{\ast} = 0.3964 m_e$ is the electron effective mass at the conduction band minimum, and $C = 0.4191$ eV$^{-1}$~\cite{Liu_ITO_2014} is the first-order non-parabolicity factor. By substituting the $\e$-$k$ relation into Eq.~\eqref{eq:eps_pump_adiab}, expressing the integrand as a power series of $q$, and converting the integral over ${\bf k}$ to an integral over $\e$ using the eDOS~\eqref{eq:DOS}, we obtain 
\begin{widetext}
\begin{align}\label{eq:lindhard_epsilon}
\varepsilon(t;\omega_\text{pump}) &= \varepsilon_\infty - \dfrac{ e^2}{\varepsilon_0 m_e^\ast}\int\dfrac{\rho_e(\e) f(\e,t;\omega_\text{pump})}{(\omega_\text{pump} + i \eta(\e,t;\omega_\text{pump}))^2}\dfrac{(1 + 8C\e(1 + C\e)/3)}{(1+2C\e)^3}d\e.
\end{align}
\end{widetext}
In the context of ITO modeling, Eq.~\eqref{eq:lindhard_epsilon} represents a rigorous and unique extension of Lindhard's formula to time-varying systems having a non-parabolic conduction band. In particular, the permittivity~\eqref{eq:lindhard_epsilon} includes the contribution from the non-thermal part of the electron distribution, instead of the thermal part only, as in all previous work. We show below that the deviation from equilibrium is negligible for pulse durations much longer than the $e$-$e$ relaxation time $\tau_{e\text{-}e}$ such as those used in~\cite{Guo_ITO_AM_2017,Guo_ITO_nanorod_natphoton,Boyd_Nat_Phot_2018,Kinsey_ENZ_OptMatExp,Exeter_Nat_Comm_2021,Exeter_Nat_Comm_2021,Sapienza_2022,de_Leon_Mexicans_2022_1}, but that it is non-negligible for pulse durations comparable or shorter than $\tau_{e\text{-}e}$. Moreover, the damping rate $\eta$ in Eq.~\eqref{eq:lindhard_epsilon} is obtained self-consistently from the electron distribution via the $e$-$e$, $e$-$ph$ and $e$-$imp$ scattering rates (see~\cite{Un-Sarkar-Sivan-LEDD-I}). This is in contrast with the phenomenological methods addressing the permittivity damping term in previous studies~\cite{Guo_ITO_nanorod_natphoton,Guo_ITO_nanorods_NC_2016,Boyd_Nat_Phot_2018,Ellenbogen-Minerbi-ITO,de_Leon_Mexicans_2022_1}. Below, we show that this self-consistent form successfully explains the experimental data~\cite{Boyd_NLO_ENZ_ITO,Xian_group_ITO_2019,Sapienza_2022} without introducing phenomenological adjustments. 

If the damping term $\eta$ is approximated to be energy-independent, then, Eq.~\eqref{eq:lindhard_epsilon} reproduces the Drude formula for the relative permittivity with an electron-distribution-dependent plasma frequency~\cite{Guo_ITO_nanorod_natphoton,Guo_ITO_nanorods_NC_2016,Boyd_Nat_Phot_2018,Ellenbogen-Minerbi-ITO,de_Leon_Mexicans_2022_1}~\footnote{In this case, the factor $1/(\omega_\text{pump} + i \eta)^2$ can be factored out of the integral, so that one can write $\varepsilon = \varepsilon_\infty - \omega_p^2/(\omega + i \eta)^2$, where
\begin{multline*}
\omega_p^2[f(\e)] = \dfrac{e^2}{\varepsilon_0 m_e^\ast}\int\rho_e(\e)f(\e)\dfrac{(1 + 8 C \e(1 + C\e)/3)}{(1 + 2 C\e)^3} d\e \\
\xlongequal[\text{by parts}]{\text{integration}} \dfrac{e^2}{3 \pi^2 \varepsilon_0 m_e^\ast} \int \left[\dfrac{2 m_e^\ast}{\hbar^2}\e(1 + C\e)\right]^{3/2}\\
(1 + 2C\e)^{-1} \left(-\dfrac{\partial f}{\partial \e}\right)d\e,
\end{multline*}
reproducing the widely-used formula for the electron temperature dependent plasma frequency~\cite{Guo_ITO_nanorod_natphoton,Guo_ITO_nanorods_NC_2016,Boyd_Nat_Phot_2018,Ellenbogen-Minerbi-ITO,de_Leon_Mexicans_2022_1} for non-parabolic band dispersion.}, namely,
\begin{multline}\label{eq:omgp_sq}
\omega_p^2[f(\e)] = \dfrac{e^2}{3 \pi^2 \varepsilon_0 m_e^\ast} \int \left[\dfrac{2m_e^\ast}{\hbar^2}\e(1 + C\e)\right]^{3/2}\\
(1 + 2C\e)^{-1} \left(-\dfrac{\partial f}{\partial \e}\right)d\e.
\end{multline}
%{\bf IW - consider moving this sentence before the discussion of the T-dependent plasma frequency - We mention the T-dependent plasma frequency when we discuss the dynamics of $\varepsilon''$ (Fig. 4(d)). Since the discussion there is already tedious, would it be better to put this sentence here?} 
The extra factor for $C \neq 0$ in the integral of Eqs.~\eqref{eq:lindhard_epsilon} and~\eqref{eq:omgp_sq} originates from the nonparabolic energy-momentum relation of ITO when expanding the denominator of Eq.~\eqref{eq:eps_pump_adiab} in a power series of $q$; it reflects the energy-dependence of the electron effective mass due to the non-parabolicity. 

Finally, if the electron distribution function can be approximated by a thermal (i.e., Fermi-Dirac) distribution, then, the (dynamics of the) permittivity can be described as a function of the electron and phonon temperatures only. Here, the electron and phonon temperatures can be obtained by solving the TTM (see Appendix~\ref{app:eTTMnTTM}).

\subsection{Heuristic explanation of temperature and permittivity dynamics}
Before dwelling into the detailed description of the temperature and permittivity dynamics, it is useful to provide an intuitive and simple description of those as a function of the illumination intensity. In particular, the maximum of the electron and phonon temperatures  and local fields can be qualitatively understood by the illumination-induced ENZ resonance shift.

The absorption of the pump pulse energy results in an increase of the electron temperature (see Fig.~\ref{fig:Te_Tph_dym}(a) below) and thus an increase in the real part of the permittivity (see Fig.~\ref{fig:epsilon_dynamics}(a) below). This causes the ENZ resonance to slip away from the incoming frequencies so that the local field and the absorptivity decrease (see Fig.~\ref{fig:normUabs}); consequently, the total absorbed energy, the maximum electron and phonon temperature increases sublinearly (rather than linearly) with the illumination intensity. As shown below, this effect explains the high damage threshold of ITO. The changes of the imaginary part of the ITO permittivity are also large, and similar in nature to those occurring in noble metals (see, e.g.,~\cite{Gurwich-Sivan-CW-nlty-metal_NP,IWU-Sivan-CW-nlty-metal_NP}), but are secondary compared to the large ENZ resonance shift.

\section{Rigorous results}\label{sec:results}
\begin{figure}[ht]
\centering
\includegraphics[width=1\columnwidth]{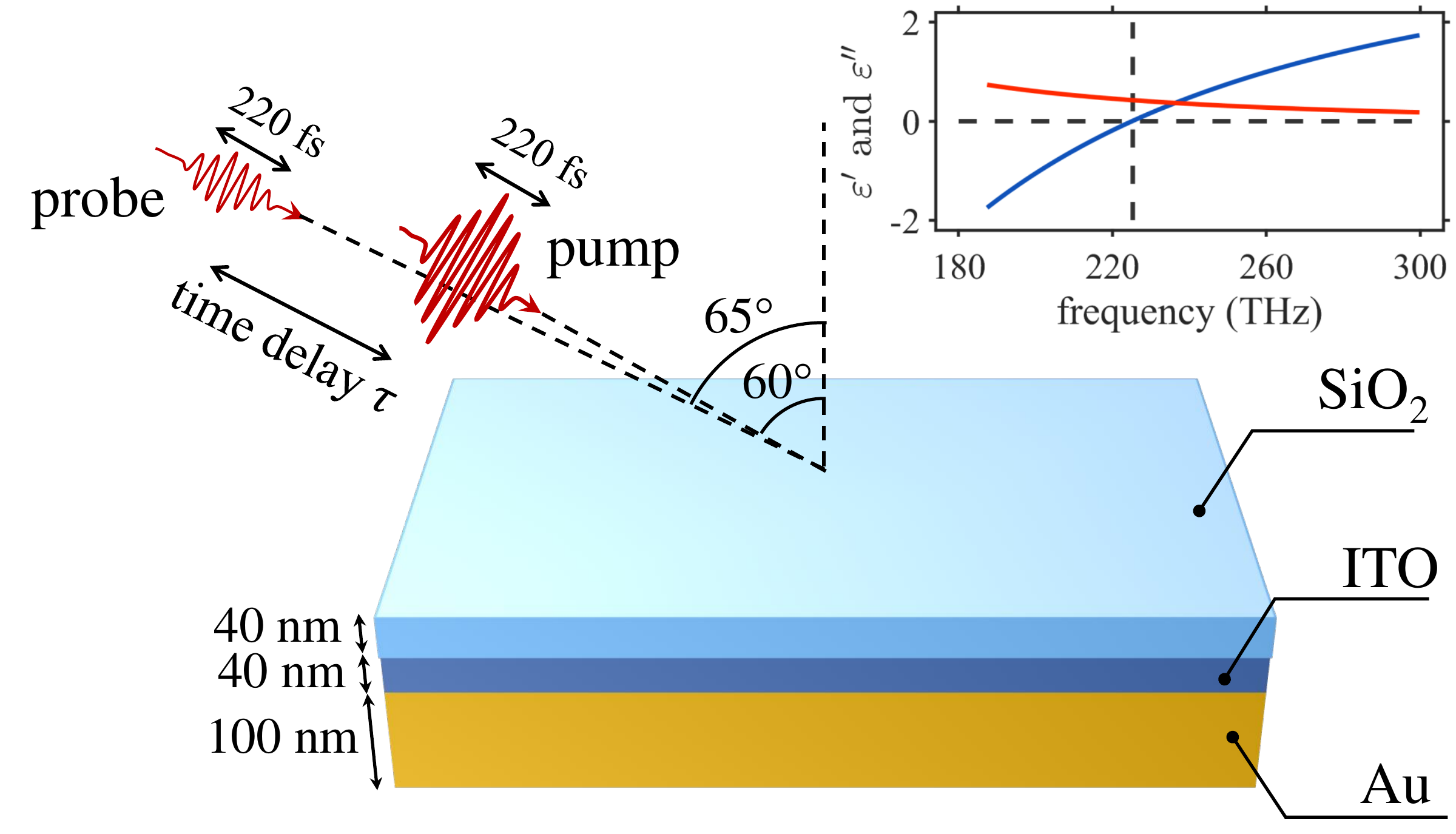}
\caption{(Color online) Schematic diagram of the setup considered here and in experiment~\cite{Sapienza_2022}. The sample consists of a 40 nm thin film of ITO deposited on glass and covered by a 100 nm gold film. A pump and a probe pulse were obliquely incident on the sample at angles of 60$^\circ$ and 65$^\circ$, respectively. The probe pulse arrives at the sample with a time delay $\tau$ after the pump pulse. {\XYZ The inset shows the real (blue solid line) and imaginary (red solid line) parts of the ITO permittivity at 300 K (i.e., plug in the Fermi-Dirac distribution at 300 K to Eq.~\eqref{eq:lindhard_epsilon}). The black dashed lines indicate the ENZ point at $\sim 225$ THz.} 
}
\label{fig:scheme}
\end{figure}

As a specific example, we consider a system similar to the one studied in~\cite{Sapienza_2022}, where a pump and a probe pulse are obliquely incident on a sample consisting of a 40 nm thin film of ITO deposited on glass and covered by a 100 nm gold film, see Fig.~\ref{fig:scheme}. {\XYZ The real part of the ITO permittivity at 300 K is near zero at $\sim 225$ THz.} This configuration is effective mainly because of the low wave impedance of the incoming pulse which enables good light penetration into the ITO.

The electric fields of the incident pump and probe pulses are written as the product of an envelope and a carrier wave. $\bm{\mathscr{E}}_\text{pump}(t) = ({\bf E}_\text{pump,0}/2) e^{- 2 \ln 2 (t/\tau_\text{pump})^2} e^{-i\omega_\text{pump}t} + \text{c.c.}$ and $\bm{\mathscr{E}}_\text{probe}(t;\tau) = ({\bf E}_\text{probe,0}/2) e^{- 2 \ln 2 ((t - \tau)/\tau_\text{probe})^2} e^{-i\omega_\text{probe}(t-\tau)} + \text{c.c.}$, where ${\bf E}_\text{pump,0}$ is related to the peak intensity by $I_0 = \dfrac{|{\bf E}_\text{pump,0}|^2}{2} \sqrt{\dfrac{\varepsilon_0}{\mu_0}}$, {\XYZ and the envelope is taken to be Gaussian~\footnote{Note that within our formulation we can take any envelope as long as the pulse duration is much longer than $1/\eta$ in Eq.~\eqref{eq:lindhard_epsilon}}.} Here $\tau_\text{pump} = \tau_\text{probe} = 220$ fs are the pulse durations, $\omega_\text{pump}/2\pi = \omega_\text{probe}/2\pi = 230$ THz are the carrier frequencies, $\tau$ is the time delay between the arrival of pump and probe pulses, and ${\bf E}_\text{probe,0}$ is the probe pulse peak field which is assumed to be weak enough so that it does not affect the electron dynamics in the ITO. Both pump and probe pulses are set to be $p$-polarized to satisfy the ENZ resonance condition. In the self-consistent calculation, we update the local field by solving  Eq.~\eqref{eq:Maxwell_eq_adiab} via the transfer matrix method~\cite{Sapienza_2022,TMM_IEEE,Principles_of_Optics}. Since the thickness of the ITO layer is much smaller than the pump wavelength, the absorbed power density is almost uniform and thus can be approximated by its spatial average for simplicity. Moreover, we assume that the incident pulse has a large spot size so that the non-uniformity of $T_e$ and $T_{ph}$ in the transverse direction can be neglected. The initial temperature of the electrons and phonons are set to be room temperature $T_0 = 300$ K. 

We show below results for pump peak intensities of $I_0 = 5$ GW/cm$^2$ to $75$ GW/cm$^2$. For simplicity, we ignore any nonlinearities in the glass and gold layers. This is justified in light of the extremely strong nonlinearity of ITO. For even higher intensities, substantial amounts of electrons would accumulate at the top edge of the conduction band. Resolving this requires one to account for spontaneous and stimulated emission of photons, photoemission of electrons, and even for higher-order non-parabolicity of the band structure, multi-photon processes etc.. The study of these effects is left to future studies. We also compare our electron dynamics model with the TTM, see details in Appendix~\ref{app:eTTMnTTM}. This provides a better understanding of the effects of the non-thermal electron distribution on the ITO nonlinearity.

\subsection{Temperature dynamics}\label{sec:T-dynamics}

\begin{figure}[b]
\centering
\includegraphics[width=\columnwidth]{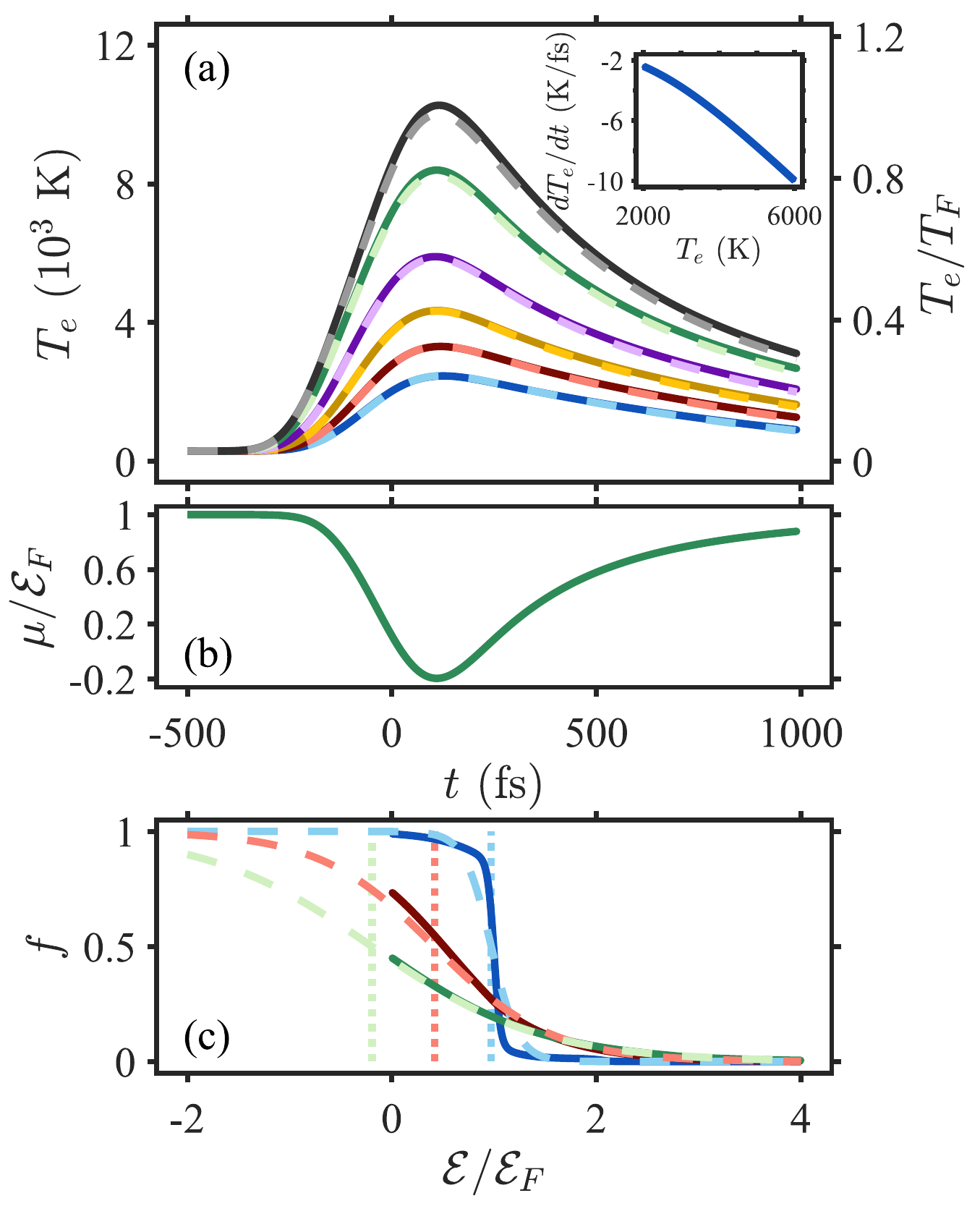}
\caption{(Color online) %{\bf IW - maybe better add a y scale with absolute temperature to (a) and to 5(a). } 
(a) Electron temperature {\XYZ (normalized to the Fermi temperature $T_F = 10204$ K on the right axis)} as a function of time for pulse intensity $I_0 = 2.5$ GW/cm$^2$ (blue lines), 5 GW/cm$^2$ (orange lines), 10 GW/cm$^2$ (yellow lines), 22 GW/cm$^2$ (purple lines), 50 GW/cm$^2$ (green lines) and 75 GW/cm$^2$ (black lines). The solid lines and the dashed lines represent the extracted temperatures from the solution of the BE (Eq.~(3) in~\cite{Un-Sarkar-Sivan-LEDD-I}) and the effective temperatures obtained from the TTM (see Appendix~\ref{app:eTTMnTTM}), respectively. The inset shows the decay rate of $T_e$ as a function of the electron temperature (the phonon temperature is set to be 300 K). (b) The (instantaneous) chemical potential (see details in~\cite{Un-Sarkar-Sivan-LEDD-I}) as a function of time for pulse intensity $I_0 = 50$ GW/cm$^2$. (c) The electron distribution (solid lines) as a function of energy at $t = -220$ fs (blue), $t = -55$ fs (red) and $t = 110$ fs (green) for the case of $I_0 = 50$ GW/cm$^2$, the same as (b). The dashed lines are the thermal distributions at the extracted electron temperatures and the vertical dashed lines represent the chemical potential corresponding to each of these lines; a slightly imperfect match is seen only at the early stages.}
\label{fig:Te_Tph_dym}
\end{figure}

For the purpose of  characterizing the solution of the ANTH$\mathrsfso{E}$M quantitatively, we extract the electron temperature (denoted as {\em extracted} $T_e$) from the electron distribution based on the total energy of the electron system~\footnote{The total energy of the electron system $\U$ is the first moment of the electron distribution, i.e., $\U = \displaystyle \int \e \rho_e(\e) f(\e) d\e$. The electron temperature associated with an electron distribution $f(\e)$ is then determined by
\begin{align*}
\int \e \rho_e(\e) f^T(\e,\mu(T_e),T_e) d\e = \int \e \rho_e(\e) f(\e) d\e,
\end{align*}
where $\mu(T_e)$ is the chemical potential given by Eq.~\eqref{eq:num_conserv_mu}, see details in~\cite{Un-Sarkar-Sivan-LEDD-I}. }. We plot the {\em extracted} electron temperature along with the {\em effective} electron temperature obtained from the TTM as a function of time for increasing illumination intensities. Fig.~\ref{fig:Te_Tph_dym}(a) shows the dynamics of a rapid (pulse duration-limited) increase followed by electron cooling familiar from noble metals~\cite{Stoll_review}. However, Fig.~\ref{fig:Te_Tph_dym}(a) also shows that, unlike noble metals, the electron subsystem reaches extremely high electron temperatures {\XYZ (in particular, $10^4$ K)}, i.e., a considerable fraction of the Fermi temperature {\XYZ ($\sim T_F$)}. Notably, since the pump pulse duration is much longer than the $e$-$e$ relaxation time $\tau_{e\text{-}e}$ (a few tens of fs, see Fig.~\ref{fig:tau_ee_ephn}(a)), the effective $T_e$ shows an excellent agreement with the dynamics of the extracted $T_e$. As a result, the $T_e$ dynamics can be understood by energy balance (Eq.~\eqref{eq:TTM_Te}), where the electron temperature decay rate ${dT_e}/{dt}$ is related to the electron heat capacity $C_e$, the electron-phonon energy coupling coefficient $G_{e\text{-}ph}$ and to the temperature difference between electrons and phonons~\footnote{The electron temperature decay rate can be obtained by setting $p_\text{abs} = 0$ in Eq.~\eqref{eq:TTM_Te}. Since the change of $T_{ph}$ is much smaller than that of $T_e$, we set $T_{ph} = 300$ K when calculating the electron temperature decay rate in the inset of Fig.~\ref{fig:Te_Tph_dym} (a).} by ${dT_e}/{dt} = - {G_{e\text{-}ph}(T_e) (T_e-T_{ph})}/{C_e(T_e)}$. {\XYZ The inset of Fig.~\ref{fig:Te_Tph_dym}(a)} shows that the electron temperature decay rate is faster for higher $T_e$, or more specifically, the slopes of the $T_e(t)$ curves become less negative as $t$ increases ($T_e$ decreases). This slower decay originates from the proportionality to the temperature difference between electrons and phonons and the simultaneous mere sublinear increase of $C_e$ with $T_e$ (see~\cite{Un-Sarkar-Sivan-LEDD-I}). Fig.~\ref{fig:Te_Tph_dym}(a) also shows that the slope of $T_e(t)$ is more negative for higher $I_0$ (higher $T_e$).

The drastic increase of $T_e$ for $I_0 \geq 50$ GW/cm$^2$ reveals a rather surprising result - the effective chemical potential $\mu$ can become negative. Specifically, Fig.~\ref{fig:Te_Tph_dym}(b) plots $\mu$ as a function of time for the case of $I_0 = 50$ GW/cm$^2$, showing that the chemical potential goes below zero after the pulse peaks at zero fs, and again recovers to positive values as $T_e$ decays. Consequently, the electron distribution is initially flatter, as seen in Fig.~\ref{fig:Te_Tph_dym}(c), and eventually becomes a distribution that is analogous to the electron distribution in a semiconductor~\cite{Sarkar-Un-Sivan-Dubi-NESS-SC}.

\begin{figure}[b]
\centering
\includegraphics[width=1\columnwidth]{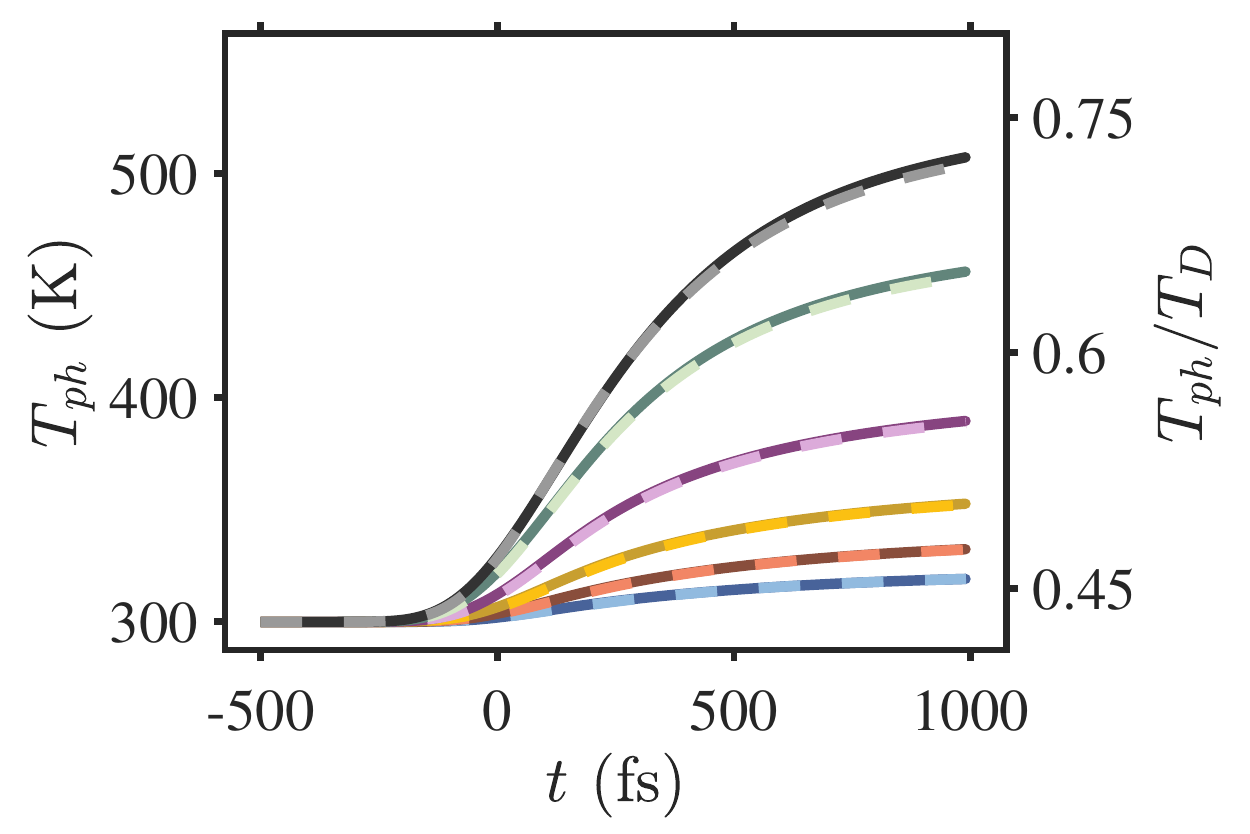}
\caption{(Color online) The same as Fig.~\ref{fig:Te_Tph_dym}(a) but for the phonon temperature $T_{ph}$ {\XYZ (normalization to the Debye temperature on right axis)}.}\label{fig:Tph_dym}
\end{figure}

The cooling of the electrons results in an increase of $T_{ph}$, {\XYZ see Fig.~\ref{fig:Tph_dym}. Fig.~\ref{fig:Tph_dym}} also shows excellent agreement between the $T_{ph}$ dynamics in the ANTH$\mathrsfso{E}$M and the TTM (Eq.~\eqref{eq:TTM}). Moreover, the phonon temperature dynamics show a decreasing growth rate {\XYZ with time.}%{\bf IW - with?}. 'This, again, can be understood by the equation of energy balance Eq.~\eqref{eq:TTM_Tph}, where the increase rate of $T_{ph}$ is given by $dT_{ph}/dt = G_{e\text{-}ph}\cdot(T_e-T_{ph})/C_{ph}$. As $t$ increases, the temperature difference between electrons and phonons becomes smaller, leading to a smaller $T_{ph}$ increase rate. With the same reasoning, the increase rate of $T_{ph}$ in ITO is much faster than that in noble metals because of the much higher $T_e$ reached~\cite{Un-Sarkar-Sivan-LEDD-I}.

\subsection{Permittivity dynamics (nonlinear optical response)}\label{subsec:epsilon-dynamics}
\begin{figure}[h]
\centering
\includegraphics[width=1\columnwidth]{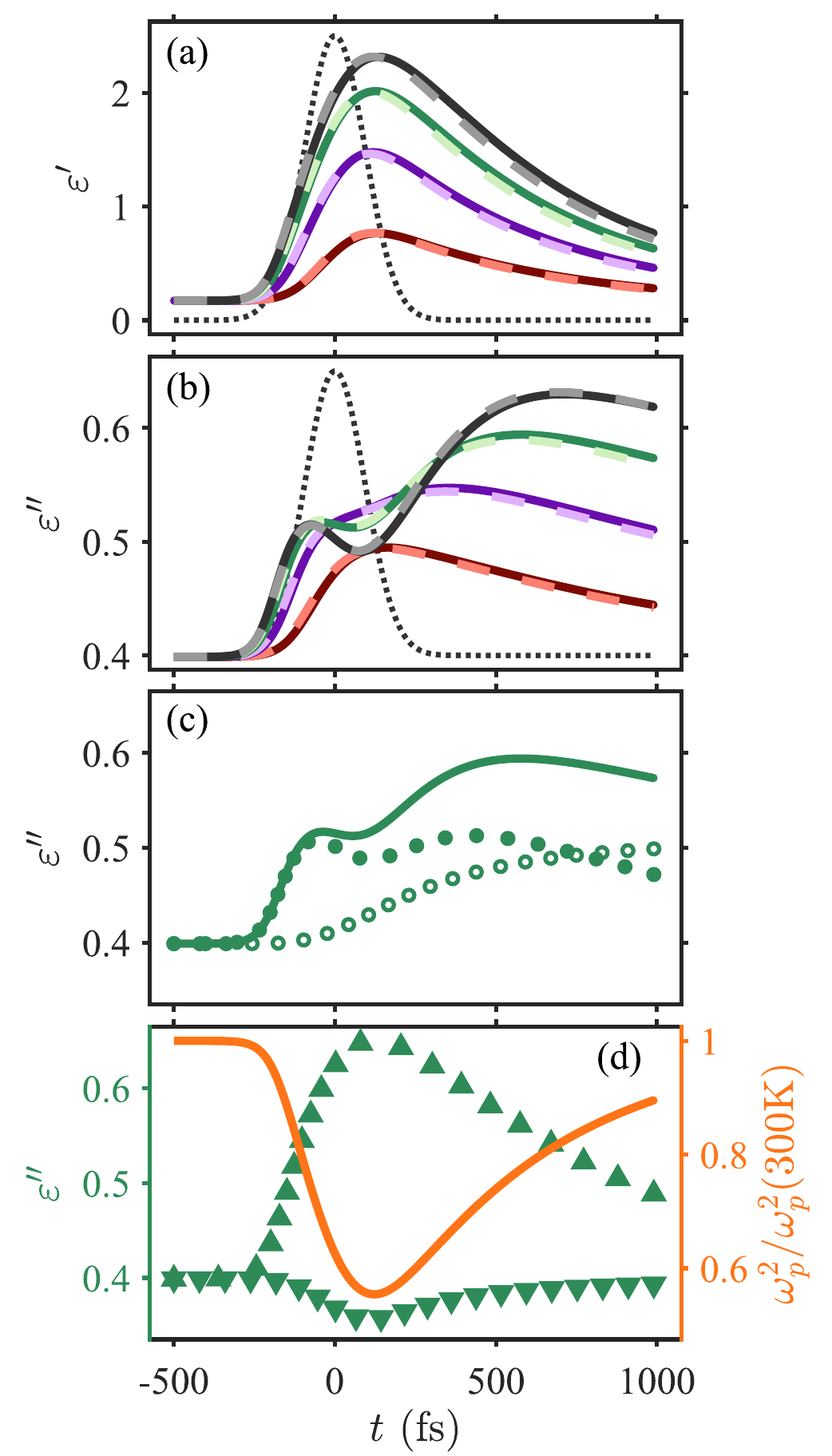}
\caption{(Color online) %{\bf IW - add pulse profile here?} 
(a) The real part and (b) the imaginary part of the permittivity~\eqref{eq:lindhard_epsilon} as a function of the time delay for the pulse intensity $I_0 = 5$ GW/cm$^2$, 22 GW/cm$^2$, 50 GW/cm$^2$ and 75 GW/cm$^2$ (the color used is the same as Fig.~\ref{fig:Te_Tph_dym}) at the carrier frequency. (c) The separation of the contributions to $\varepsilon''$ due to the change of $\eta_{e\text{-}ph}$ (open dots); and $T_e$ (filled dots) for $I_0 = 50$ GW/cm$^2$. The solid line is the same as in (b). (d) The separation of the effects of $T_e$ on $\varepsilon''$ via the change of $\eta_{e\text{-}e}$ (up-pointing triangles) and $\omega_p^2$ (down-pointing triangles). The orange solid line shows the change of $\omega_p^2$ (normalized to its value at 300 K) as a function of the time delay. {\XYZ The thin black dotted lines in (a) and (b) represent the pump pulse intensity profile.}
}
\label{fig:epsilon_dynamics}
\end{figure}

In Fig.~\ref{fig:epsilon_dynamics} we plot the dynamics of the ITO permittivity based on the solution of the ANTH$\mathrsfso{E}$M. The result shows a remarkable match to a permittivity calculation that relies on pure thermal electron distributions with the effective $T_e$. This match enables a simple interpretation of the permittivity in terms of the dynamics of the temperatures shown in Fig.~\ref{fig:Te_Tph_dym}(a) and (b). This excellent match is remarkable since it does not require any of the phenomenological corrections introduced in~\cite{de_Leon_Mexicans_2022_1}. In the latter, the permittivity calculations matched the experimental reflectance/transmittance~\cite{Boyd_NLO_ENZ_ITO} only if the damping coefficient $\eta$ was assumed to increase linearly with $T_e$, the $e$-$ph$ coupling coefficient was reduced by a factor of 8, and the $e$-$ph$ energy transfer from the non-thermal part of the electron distribution is assumed to be negligible. The self-consistent treatment presented here not only naturally connects the damping $\eta$ with $T_e$ and $T_{ph}$, but also explains the experimentally-observed increase of the imaginary part of the permittivity induced by the illumination~\cite{Xian_group_ITO_2019}. 

The most striking aspect of the dynamics is the drastic change in the real part $\varepsilon'$ (a 12-fold increase for $I_0 = 50$ GW/cm$^2$), which is a result of the initial proximity to the ENZ point~\cite{Boyd_NLO_ENZ_ITO,Shalaev_Faccio_NLO_ENZ,Exeter_Nat_Comm_2021,Sapienza_2022}. One can see that the dynamics of $\varepsilon'$ shares similar features with the time evolution of $T_e$, (compare Figs.~\ref{fig:Te_Tph_dym}(a) and~\ref{fig:epsilon_dynamics}(a)). This is caused by the (nearly) linear decrease of the effective plasma frequency with $T_e$ due to the non-parabolicity, which has been explained in many previous studies~\cite{Guo_ITO_nanorod_natphoton,Yang_polarization_switching_NP_2017,Boyd_Nat_Phot_2018,Exeter_Nat_Comm_2021}. Therefore, by Fig.~\ref{fig:Te_Tph_dym}(a), the temporal decay of the ITO permittivity is faster for higher effective $T_e$. This indicates that the permittivity dynamics shows the opposite behaviour to the experimentally-observed slowing of the reflection decay at high intensities in~\cite{Sapienza_2022}. We return to this issue in Section~\ref{subsec:probe_dynamics}. The positive correlation between $\varepsilon'$ and the effective $T_e$ also explains why the maximum change in $\varepsilon'$ occurs later than the peak of the pump pulse. This is because in the early stages the change of effective $T_e$ is approximately proportional to the total absorbed energy density, i.e., $\Delta \varepsilon'(t) \sim \Delta T_e(t) \sim \displaystyle\int^t_{-\infty} p_\text{abs}(t')dt'$.

In that regard, although ITO is a Drude material, the (large than realized before) temperature-induced change in the real part of the permittivity makes the ENZ nonlinear optical response of ITO qualitatively different from that of noble metals; indeed, in the latter, the light-induced changes to the imaginary part dominate over the changes to the real part~\cite{delFatti_nonequilib_2000,valle_transient_nlty_response_2015}. Moreover, at the ENZ point, not only the field enhancement and absorption are strong, but also the relative change $\varepsilon'$ ($=\Delta\varepsilon'/\varepsilon'$) and thus the nonlinear optical response is maximized. 

By contrast, $\varepsilon''$ shows a more complicated dynamics because it is proportional to the product of $\omega_p^2$ (Eq.~\eqref{eq:omgp_sq}) and the damping term $\eta$, and thus depends on both $T_e$ and $T_{ph}$. To see this, in Fig.~\ref{fig:epsilon_dynamics}(c), we separate the contribution due to the change of $T_e$ (by setting $T_{ph} = 300$ K) and of $T_{ph}$ (setting $T_e = 300$ K). Since $T_{ph}$ affects $\tau_{e\text{-}ph}^{-1}$ linearly~\cite{Un-Sarkar-Sivan-LEDD-I} but does not affect $\tau_{e\text{-}e}^{-1}$ nor $\omega_p$ (see Eq.~\eqref{eq:omgp_sq}), the contribution due to $T_{ph}$ results in a slow increase of $\varepsilon''$ (see open dots in Fig.~\ref{fig:epsilon_dynamics}(c)). Fig.~\ref{fig:epsilon_dynamics}(c) also shows that the complicated dynamics of $\varepsilon''$ originates from the dependence on $T_e$ (see solid dots in Fig.~\ref{fig:epsilon_dynamics}(c)). In particular, when $T_e$ increases {\XYZ within the duration of $-200\,{\rm fs} < t < 0$}, %{\bf IW -  $-200 < t < 0$?}, 
$\tau_{e\text{-}e}^{-1}$ increases but $\omega_p$ decreases, see their opposite effects on the $\varepsilon''$ dynamics separated in Fig.~\ref{fig:epsilon_dynamics}(d). Thus, the $T_e$-dependence of $\varepsilon''$ can be roughly expressed as $\varepsilon'' \sim T_e^2(1 - \alpha T_e)$~\footnote{Here, $\alpha \sim -\left.\dfrac{\partial \omega_p^2}{\partial T_e}\right|_{T_e=T_0}> 0$ represents the decrease of the effective plasma frequency with $T_e$.}, indicating that $\varepsilon''$ increases for modest electron temperatures but decreases at higher ones. Therefore, when the pump peak intensity is relatively high ($> 50$ GW/cm$^2$ in Fig.~\ref{fig:epsilon_dynamics}(b) and (c)), the standard $\varepsilon''$ dynamics of a rapid (pulse duration-limited) increase followed by electron cooling and phonon heating is accompanied by an additional drop once high electron temperatures are reached (see the green solid lines in Figs.~\ref{fig:epsilon_dynamics}(b) and (c)). That drop in $\varepsilon''$ is more pronounced as the illumination intensity is increased, compare the purple and green lines in Fig.~\ref{fig:epsilon_dynamics}(b). Importantly, the increase of the imaginary part of the ITO permittivity with time (and intensity) further indicates that the physics underlying the thermo-optic response of LEDD materials is similar to that of Drude metals (see~\cite{Gurwich-Sivan-CW-nlty-metal_NP,IWU-Sivan-CW-nlty-metal_NP}, and is the opposite to that of saturable absorbers (for which the imaginary part of the permittivity decreases upon illumination). 

\subsection{Intensity dependence and energy partition}\label{subsec:eng_partition}
\begin{figure}[t]
\centering
\includegraphics[width=0.95\columnwidth]{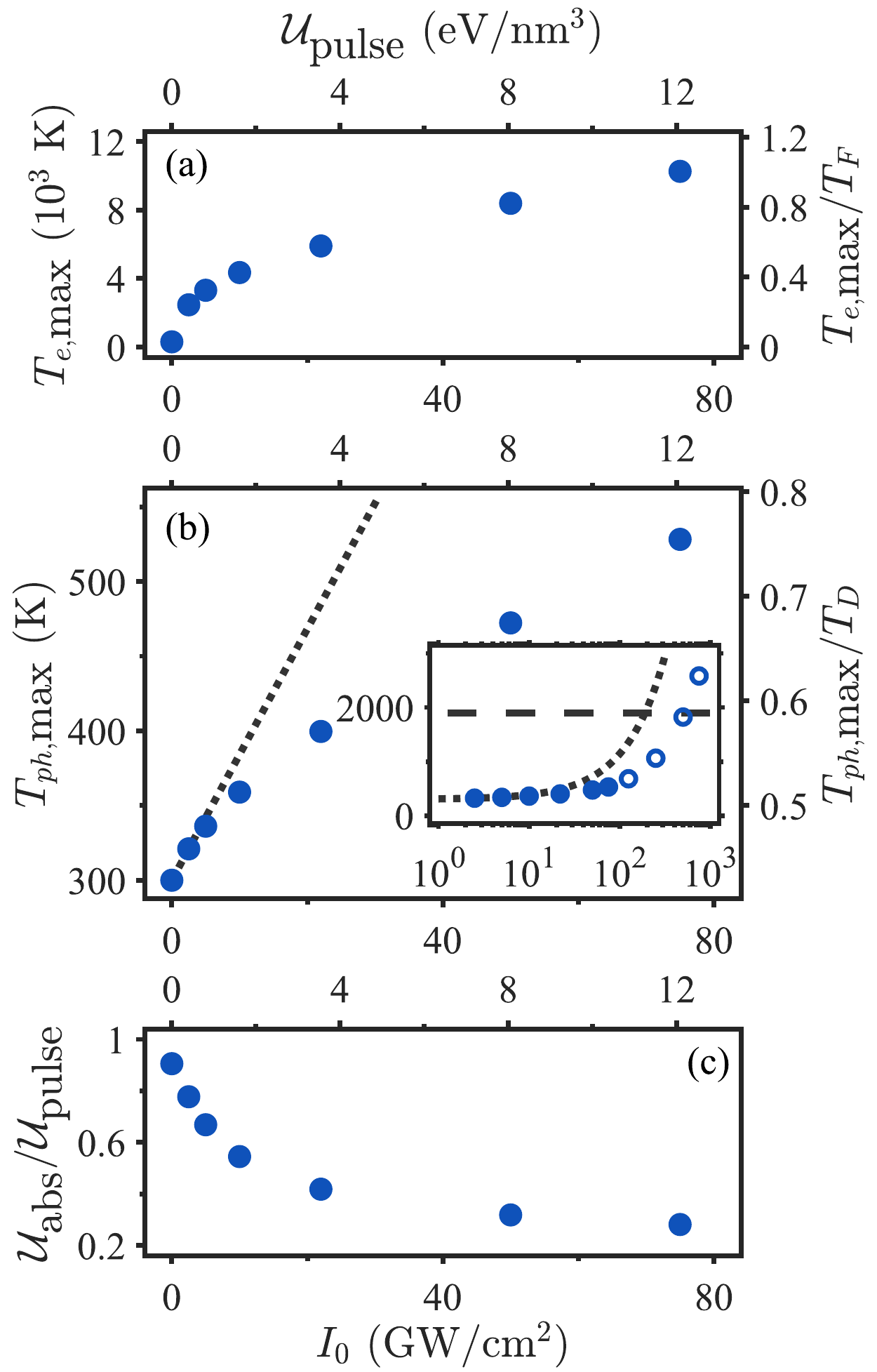}
\caption{(Color online) (a) The maximum of the electron temperature {\XYZ (normalized to the Fermi temperature on right axis)} as a function of the pulse peak intensity (extracted from Fig.~\ref{fig:Te_Tph_dym}(a)). (b) The same as (a) but for the phonon temperature $T_{ph}$ {\XYZ (normalization to the Debye temperature on right axis)}. The (filled) dots represent the data obtained from the simulation and the dotted black line represents the maximum phonon temperature obtained from an intensity-independent permittivity approximation. The inset of (b) shows the same data on a semilogx scale and is extended to 750 GW/cm$^2$. The open dots represent the estimated upper bound of the maximum $T_{ph}$ for $I_0$ up to 750 GW/cm$^2$ assuming that the absorptivity is constant for $I_0 > 75$ GW/cm$^2$. The dashed black line represents the damage threshold temperature (1900 K) of ITO reported in~\cite{Sapienza_2022}. (c) The total absorbed pump pulse energy (normalized to the pump pulse energy) as a function of the pulse peak intensity (bottom x axis)/pump pulse energy density (top x axis).}
\label{fig:normUabs}
\end{figure}

One of the direct consequence of the drastic change in $\varepsilon'$ induced by the strong illumination is the sublinear growth of the maximum of the electron and phonon temperatures with the pump peak intensity, as shown in Fig.~\ref{fig:normUabs}(a)-(b). The maximal values of $T_{ph}$, although can be estimated by extrapolating the $T_{ph}$ dynamics in Fig.~\ref{fig:Te_Tph_dym}(b), are more easily deduced from the total absorbed energy {\XYZ density} $\left(\mathcal{U}_\text{abs}(I_0) = \displaystyle\int p_\text{abs}(t,I_0) dt\right)$ from the self-consistent simulations via $T_{ph,\text{max}}(I_0) = T_0 + {\U_\text{abs}(I_0)}/{C_{ph}}$. This is based on the fact that almost all the absorbed energy transfers to the phonon subsystem before leaking to out to the surrounding~\footnote{This is because electrons stop transferring their energy once $T_e$ and $T_{ph}$ are the same. At this stage, $T_e$ is higher than its initial temperature $T_0$. This means that a small amount of the absorbed energy remains in the electron subsystem (this amount of the energy is small since $C_e\ll C_{ph}$). This excess energy will be transferred to the environment on the slower timescale neglected in this work. }. The sublinear growth of the maximum phonon temperatures with the pump peak intensity can then be understood by the absorptivity (the ratio of the total energy absorbed by ITO and the incident pump pulse energy, denoted as $\mathcal{U}_\text{pulse}$) as shown in Fig.~\ref{fig:normUabs}(c). It shows that at low illumination intensities, for which the permittivity is close to the ENZ point (see Fig.~\ref{fig:epsilon_dynamics}), most energy is absorbed (ENZ resonance). However, as the illumination intensity grows and the real part of the ITO permittivity drastically increases (see Fig.~\ref{fig:epsilon_dynamics}(a)), the resulting shift of the ENZ point causes the absorptivity to drop rapidly and the fraction of the incident energy reflected to grow. This effect can be captured only by a self-consistent simulation. This confirms that the decrease of the absorptivity does not originate from saturable absorption. 

In Fig.~\ref{fig:normUabs}(b), we further extrapolate the data obtained from the simulations up to 750 GW/cm$^2$ (the open dots) by assuming that the absorptivity remains the same for $I_0 > 75$ GW/cm$^2$, namely, $T_{ph,\text{max}}(I_0>75\ \text{GW/cm}^2) = T_0 + \dfrac{\U_\text{abs}(I_0 = 75\ \text{GW/cm}^2)}{C_{ph}}\dfrac{I_0}{75\ \text{GW/cm}^2}$. The open dots in Fig.~\ref{fig:normUabs}(b) indicates that the resulting (somewhat overestimated) $T_{ph}$ reaches the melting point of ITO ($\sim 1900$ K) for $I_0 > 500$ GW/cm$^2$. This explains the high damage threshold of ITO observed in~\cite{Sapienza_2022}.%experimentally in~\cite{Sapienza_2022}.

Finally, we note that the sublinearity of maximal $T_e$ vs. $I_0$ is more pronounced than that of $T_{ph}$. This is because, in addition to the decrease of absorptivity with $I_0$, the electron heat capacity increases with $T_e$ such that more energy is required to change $T_e$ at higher $T_e$, yielding a smaller growth of maximal $T_e$ with the peak intensity.

\subsection{Probe pulse dynamics}\label{subsec:probe_dynamics}
\begin{figure}[b]
\centering
\includegraphics[width=0.7\columnwidth]{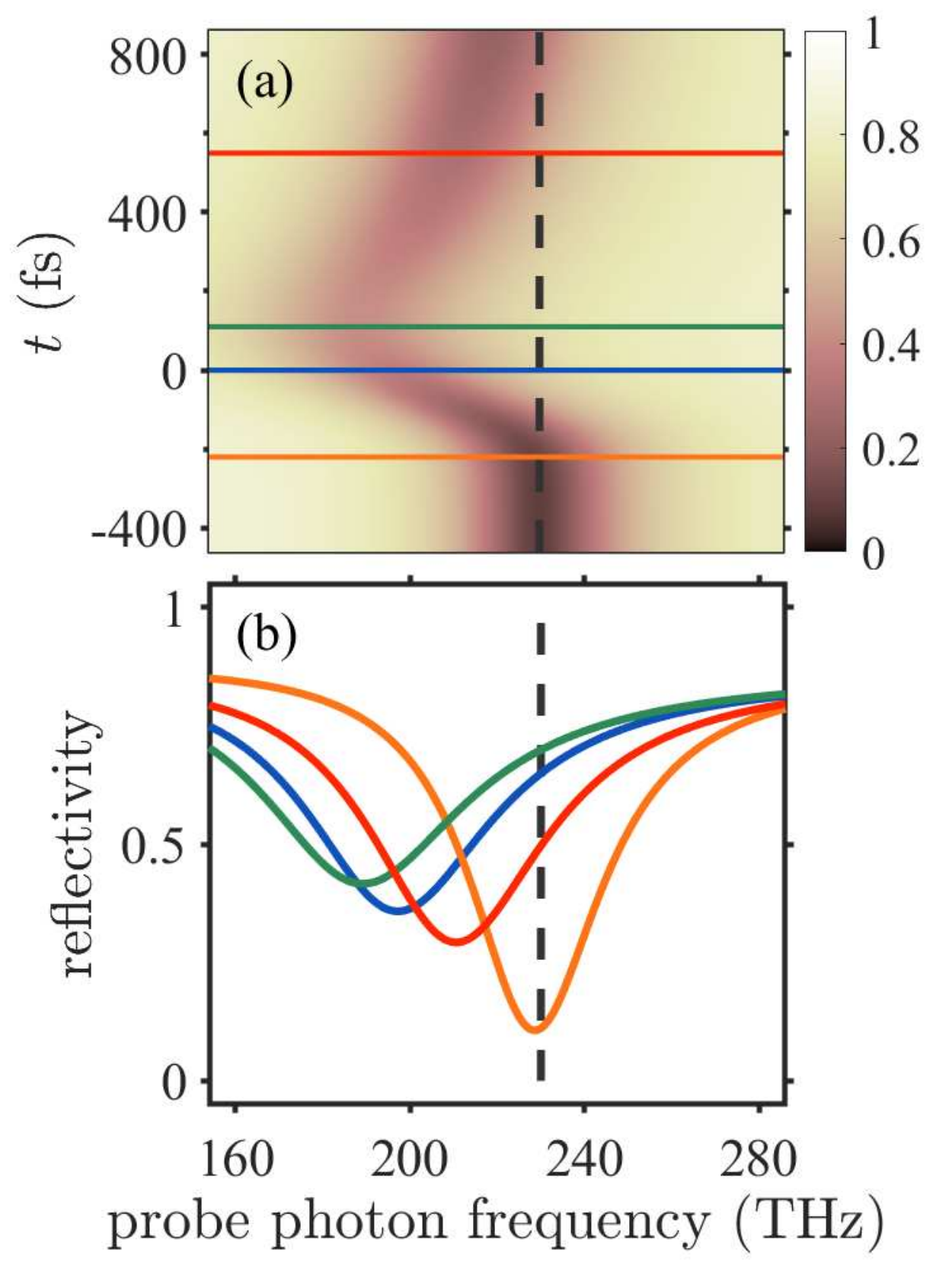}
\caption{(Color online) (a) The map of the reflectivity {\XYZ (the absolute square of the Fresnel reflection coefficient)} for the pump peak intensities of 22 GW/cm$^2$, respectively. (b) The reflectivity of the probe pulse at time delay sections $\tau =$ -220 (orange), 0 (blue), 110 (green) and 550 fs (red) in (a) and (c), respectively. The black dashed line represents the center frequency (230 THz) of the probe pulse. %{\bf IW - why not extend the red side of these plots?}
}
\label{fig:probe_reflectivity_220fs_22GW}
\end{figure}
In order to connect the ANTH$\mathrsfso{E}$M to the experimental data~\cite{Sapienza_2022}, we now calculate the temporal reflectivity and the total reflection of the probe pulse by the sample in the presence of the pump pulse. Since the probe pulse satisfies the adiabatic condition (see Section~\ref{sec:scf_epsilon}), the Amp\`{e}re's law and the Maxwell–Faraday equation for the probe pulse become
\begin{align}\label{eq:Maxwell_eq_probe_adiab}
\begin{cases}
\begin{aligned}
&\nabla\times{\bf H}_\text{probe}({\bf r},t;\tau) =  - i \omega_\text{probe} \varepsilon_0\\
&\qquad\qquad\qquad\varepsilon(t;\omega_\text{pump},\omega_\text{probe}){\bf E}_\text{probe}({\bf r},t;\tau)
\end{aligned}
\\
\nabla\times{\bf E}_\text{probe}({\bf r},t;\tau) = - i \omega_\text{probe} \mu_0 {\bf H}_\text{probe}({\bf r},t;\tau)
\end{cases},
\end{align}
where $\varepsilon(t;\omega_\text{pump},\omega_\text{probe})$ is the time-dependent permittivity for the probe pulse, namely,
\begin{widetext}
\begin{align}\label{eq:eps_probe_adiab}
\varepsilon(t;\omega_\text{pump},\omega_\text{probe}) = \varepsilon_\infty + \lim_{{\bf q} \rightarrow 0}\dfrac{2e^2}{\varepsilon_0 q^2}\int\dfrac{d^3k}{(2\pi)^3}\dfrac{f_{{\bf k} + {\bf q}}(t;\omega_\text{pump}) - f_{{\bf k}}(t;\omega_\text{pump})}{\e_{{\bf k} + {\bf q}} - \e_{{\bf k}} - \hbar\omega_\text{probe} - i \hbar(\eta_{{\bf k} + {\bf q}}(t;\omega_\text{pump}) + \eta_{{\bf k}}(t;\omega_\text{pump}))/2}.
\end{align}
\end{widetext}
Since the intensity of the probe pulse is weak enough so that it does not induce significant changes to the electron distribution, the electron distribution in Eq.~\eqref{eq:eps_probe_adiab} is the self-consistent solution obtained from Eqs.~\eqref{eq:f_neq_dynamics},~\eqref{eq:Maxwell_eq_adiab} and~\eqref{eq:eps_pump_adiab}. Again, the time variable $t$ in Eq.~\eqref{eq:Maxwell_eq_probe_adiab} and Eq.~\eqref{eq:eps_probe_adiab} is correlated with the pump pulse envelope but is not the Fourier conjugate of any photon frequency (see Section~\ref{sec:scf_epsilon}). %{\bf IW - the following is based on adiabaticity?} 
{\XYZ Based on the adiabaticity,} the temporal profile of the reflected probe pulse envelope ${\bf E}_\text{probe,ref}(t;\tau)$ is thus equal to the product of the Fresnel reflection coefficient and the incident probe pulse envelope, i.e., ${\bf E}_\text{probe,ref}(t;\tau)=r(t;\omega_\text{probe})({\bf E}_\text{probe,0}/2)e^{-2\ln2((t-\tau)/\tau_\text{probe})^2}$. %is the envelope of the reflected probe pulse.%$\bm{\mathscr{E}}_\text{probe,ref}(t;\tau) = r(t,\omega_\text{probe})({\bf E}_\text{probe,0}/2)e^{-2\ln 2((t-\tau)/\tau_\text{probe})^2}e^{-i\omega_\text{probe}(t-\tau)} + \text{c.c.}$. 
Notice that the Fresnel reflection coefficient is {\em independent} of the time-delay parameter since it originates from the (pump-induced) electron dynamics; {\XYZ its absolute square (i.e., the reflectivity) plotted in Fig.~\ref{fig:probe_reflectivity_220fs_22GW} is thus different from the spectrum of the reflected probe pulse (see below)~\footnote{Therefore, one should not compare Fig.~\ref{fig:probe_reflectivity_220fs_22GW}(a) with Fig. 3 in~\cite{Sapienza_2022}.}.} Fig.~\ref{fig:probe_reflectivity_220fs_22GW} shows that the drastic change of the real part of the permittivity induced by the pump pulse results in a shift of the ENZ resonance frequency and that this shift increases with the pump peak intensity.

\begin{figure*}[ht]
\centering
\includegraphics[width=1\textwidth]{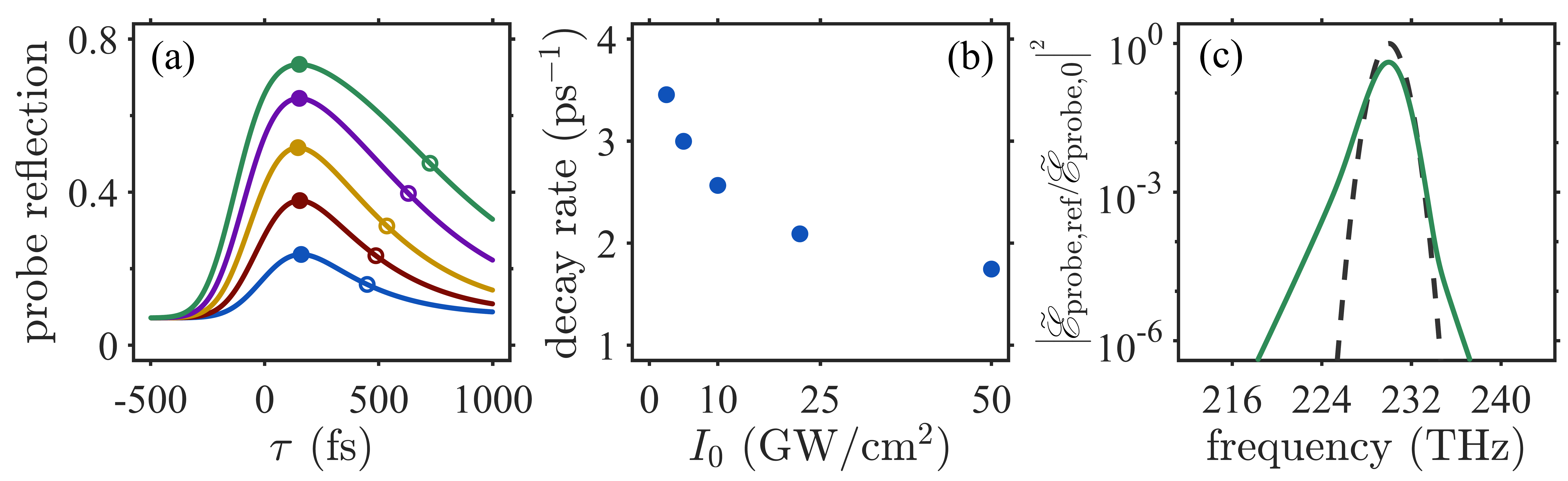}
\caption{(Color online) (a) The reflection of the probe pulse {(\XYZ the ratio of the total energy of the reflected pulse and that of the incident probe pulse)} as a function of the pump-probe delay for different pump pulse peak intensities (the same colors are used as in Fig.~\ref{fig:epsilon_dynamics}). The filled and open dots label the maximum and half-maximum of the reflection, respectively. (b) The decay rate (the inverse of the time period corresponding to the change of the reflection from its maximum to half this maximum) as a function of the pulse peak intensity. {\XYZ (c) The spectrum of the reflected probe pulse at the time delay of -90 fs (green solid line) and -500 fs (black dashed line) for the pump pulse peak intensity of 50 GW/cm$^2$.}}
\label{fig:probe_refl_decay}
\end{figure*}

The reflection of the probe pulse is then the ratio of the total energy of the reflected pulse and that of the incident probe pulse, $R(\tau) = \dfrac{\displaystyle\int |{\bf E}_\text{probe,ref}(t;\tau)|^2 dt}{\displaystyle\int |{\bf E}_\text{probe,inc}(t;\tau)|^2 dt}$; it is, therefore, a function of the time delay. Fig.~\ref{fig:probe_refl_decay}(a) shows that the reflection of the probe pulse gradually increases with the pump peak intensity, even though at a decreasing rate. In addition, the time in which the change of the reflection reduces from its maximum to half of this maximum is longer for higher pump peak intensities, see Fig.~\ref{fig:probe_refl_decay}(b). This finding matches the observations in~\cite{Sapienza_2022}. The reason for that is that when the pump peak intensity is high enough, the induced frequency shift of the ENZ resonance is larger than its resonance width (see Fig.~\ref{fig:probe_reflectivity_220fs_22GW}(b)) so that the dependence of the reflection change on the pump peak intensity becomes sublinear.

%\onecolumngrid
%\begin{center}
%\end{center}
%\twocolumngrid

{\XYZ The lack of time translational invariance prevents us from writing the spectrum of the reflected probe pulse as a product of the Fresnel reflection coefficient and the Fourier transform of the incident probe pulse. Therefore, we calculate the spectrum of the reflected probe pulse directly from its Fourier transform, namely, 
%The spectrum of the reflected probe pulse can be obtained from its Fourier transform, namely, 
$\tilde{\bm{\mathscr{E}}}_\text{probe,ref}(\omega;\tau) = \displaystyle\int \bm{\mathscr{E}}_\text{probe,ref}(t;\tau)e^{i\omega t}dt$. Here, $\bm{\mathscr{E}}_\text{probe,ref}(t;\tau) = {\bf E}_\text{probe,ref}(t;\tau) e^{-i\omega_\text{probe} t} + \text{c.c.} = r(t;\omega_\text{probe})({\bf E}_\text{probe,0}/2)e^{-2\ln 2((t-\tau)/\tau_\text{probe})^2 -i\omega_\text{probe} (t-\tau)}  + \text{c.c.}$ is the electric field of the reflected probe pulse. The adiabaticity ensures that the Fresnel coefficient, evaluated only at the center frequency of the probe pulse (i.e., $r(t,\omega_\text{probe})$, where $\omega_\text{probe}$ corresponds to the black dashed line in Fig.~\ref{fig:probe_reflectivity_220fs_22GW}(a)), is enough to calculate $\tilde{\bm{\mathscr{E}}}_\text{probe,ref}(\omega;\tau)$. Hence, the significant ENZ resonance frequency shift does not appear in the spectrum of the reflected probe pulse, see Fig.~\ref{fig:probe_refl_decay} (c). 
%We also note that because of the lack of time translation invariance, $\tilde{\bm{\mathscr{E}}}_\text{probe,ref}(\omega;\tau)$ is not equal to the product of the Fresnel reflection coefficient and %$\tilde{\bm{\mathscr{E}}}_\text{probe,inc}(\omega;\tau)$, where $\tilde{\bm{\mathscr{E}}}_\text{probe,inc}(\omega;\tau) = \displaystyle\int \bm{\mathscr{E}}_\text{probe,inc}(t;\tau)e^{i\omega t}dt$ is the Fourier transform of the incident probe pulse. 
%This is quite unlike the conventional wisdom of spectrum of the reflected probe pulse being the product of the Fresnel reflection coefficient and the Fourier transform of the incident probe pulse.
Moreover, due to the transient change of the reflectivity induced by the pump pulse new frequencies are generated beyond the bandwidth of the spectrum of the incident probe pulse, as shown in Fig.~\ref{fig:probe_refl_decay} (c). %However, 
Since the rising time of $\varepsilon'$ (see Fig.~\ref{fig:epsilon_dynamics}(a)) is long compared to the periodicity of the probe pulse $2\pi/\omega_\text{probe}$, the widening of the reflected probe pulse spectrum is rather weak (see the comparison between green solid and black dashed lines in Fig.~\ref{fig:probe_refl_decay}(c) at the respective FWHM). These results, therefore, agree with the observations in~\cite{Sapienza_2022}.}

\subsection{The response to shorter pulses}\label{subsec:short_pulse}
Since the duration of the pump pulses considered in previous subsections and in the experiments~\cite{Guo_ITO_AM_2017,Guo_ITO_nanorod_natphoton,Boyd_Nat_Phot_2018,Kinsey_ENZ_OptMatExp,Exeter_Nat_Comm_2021,Exeter_Nat_Comm_2021,Sapienza_2022,de_Leon_Mexicans_2022_1} are much longer than the $e$-$e$ relaxation time, the dynamics of the temperatures and of the permittivity can be well approximated by the TTM, see also the interpretation of the experimental results~\cite{Boyd_NLO_ENZ_ITO} in~\cite{de_Leon_Mexicans_2022_1}. To go beyond that description, we consider a pump pulse with a duration of 30 fs which is comparable with the $e$-$e$ relaxation time but still satisfies the condition for the adiabatic approximation, so that we can employ the model described in Section~\ref{sec:formulation} to calculate the electronic and optical responses. To make a fair comparison, the peak intensity of the 30 fs pulse is chosen to be 161 GW/cm$^2$ so that the pulse energy is the same as the 220 fs pulse with $I_0 = 22$ GW/cm$^2$ used so far. We further compare the results based on ANTH$\mathrsfso{E}$M with the TTM as well as on the extended TTM (eTTM), see details in Appendix~\ref{app:eTTMnTTM}. The comparison above therefore allows us to pinpoint the role of the non-thermal electrons on the nonlinear response of ITO. %We also discuss the validity of using the third-order nonlinear optical susceptibility to describe the ultrafast optical nonlinearity.

%\onecolumngrid
%\begin{center}
\begin{figure*}[ht]
\centering
\includegraphics[width=\textwidth]{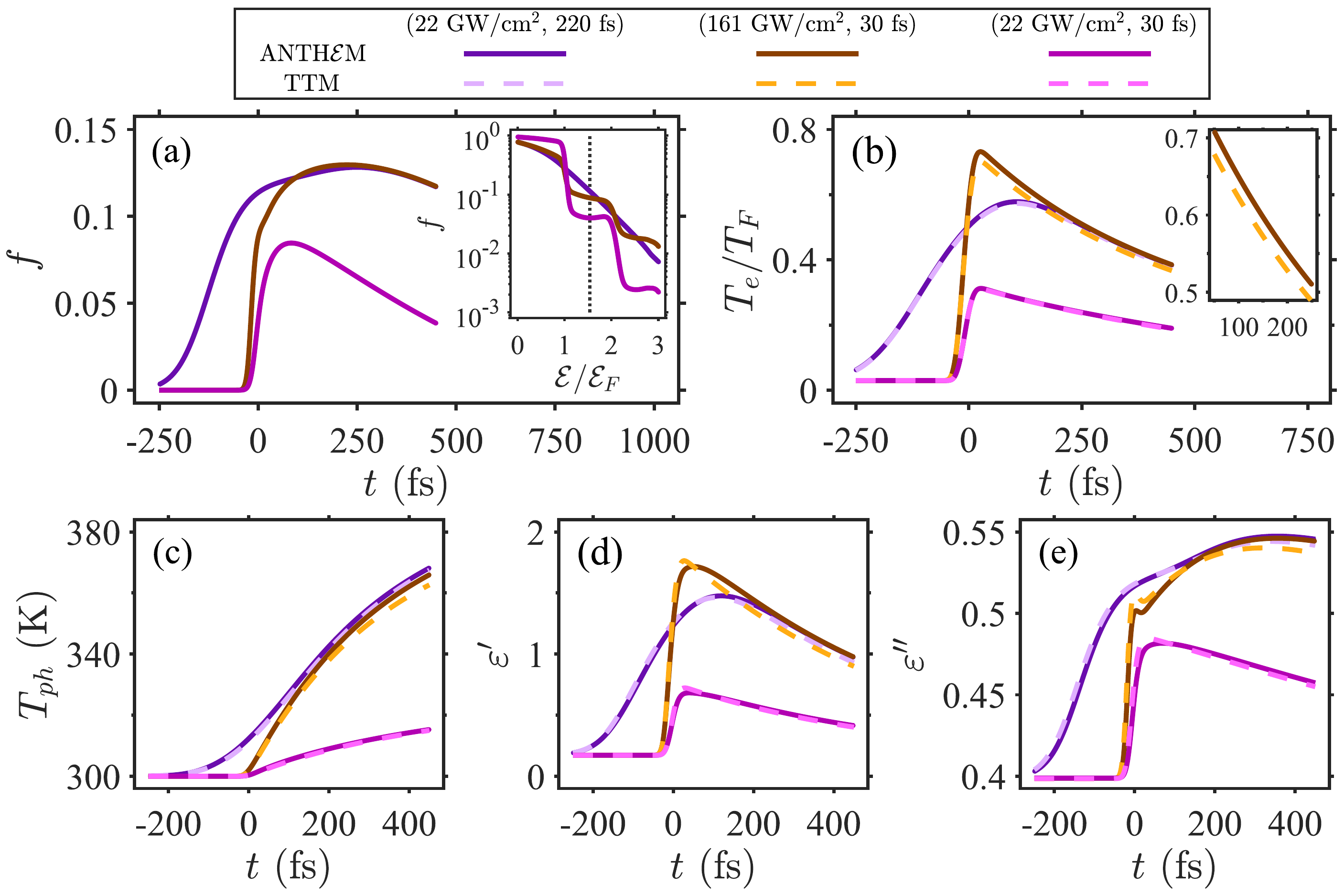}
\caption{(Color online) The time dependence of (a) the electron distribution at $\e = \e_F + \hbar\omega_\text{pump}/2$, (b) the electron temperature {\XYZ (normalized to the Fermi temperature)}, (c) the phonon temperature {\XYZ (normalized to the Debye temperature)}, (d) the real part and (e) the imaginary part of the ITO permittivity. The solid and dashed lines in (b) and (c) are obtained from the ANTH$\mathrsfso{E}$M and from the TTM, respectively. In (d) and (e), the solid and dashed lines are calculated from Eq.~\eqref{eq:eps_pump_adiab} using the electron distribution obtained from the ANTH$\mathrsfso{E}$M and the thermal distribution with the effective $T_e$, respectively. The purple, brown/orange and magenta lines respectively represent the cases of 22 GW/cm$^2$ and 220 fs, 161 GW/cm$^2$ and 30 fs, and 22 GW/cm$^2$ and 30 fs. The inset in (a) shows the electron distribution at $t = 0$ fs. The black dotted line labels the energy $\e = \e_F + \hbar\omega_\text{pump}/2$. The inset in (b) is the zoom-in at $t = $ 50 - 250 fs, showing the difference in $T_e$ between the ANTH$\mathrsfso{E}$M and TTM ($\sim 200$ K) more clearly.}
\label{fig:fe_Te_Tphn_epsm_220_vs_30}
\end{figure*}
%\end{center}
%\twocolumngrid

Fig.~\ref{fig:fe_Te_Tphn_epsm_220_vs_30} shows that the dynamics of the electron distribution, extracted electron temperature and permittivity for the case of the 30 fs pulse with peak intensity 161 GW/cm$^2$ (brown solid lines) are qualitatively the same as the case of 220 fs pulse with peak intensity 22 GW/cm$^2$ (purple solid lines) studied in Figs.~\ref{fig:Te_Tph_dym} and~\ref{fig:epsilon_dynamics}, except for moderate quantitative differences in the early stages. Specifically, after the pump pulse hits the sample, the extracted electron temperature and the permittivity for the case of 30 fs pulse with $I_0 = 161$ GW/cm$^2$ increase rapidly on the time scale of the pulse duration, much faster than that for the case of 220 fs pulse with $I_0 = 22$ GW/cm$^2$ (see Fig.~\ref{fig:fe_Te_Tphn_epsm_220_vs_30}(a), (b), (d) and (e)), {\XYZ resulting in a similar probe pulse reflection decay and a much wider reflected pulse spectrum (see Fig.~\ref{fig:probe_refl_decay_widen_v2_220_vs_30} in Appendix~\ref{app:probe_dyn_short_pulse}).} In addition, comparing with the case of 220 fs pulse with $I_0 = 22$ GW/cm$^2$, during the much shorter $T_e$ rise time there is much less energy transfer from the electron to the phonon subsystem, resulting in a higher maximal value of the electron temperature (Fig.~\ref{fig:fe_Te_Tphn_epsm_220_vs_30}(b)), and thus, in a larger change of $\varepsilon'$ (Fig.~\ref{fig:fe_Te_Tphn_epsm_220_vs_30}(d)), and a more pronounced drop in the dynamics of $\varepsilon''$ (Fig.~\ref{fig:fe_Te_Tphn_epsm_220_vs_30}(e)). 

Moreover, Fig.~\ref{fig:fe_Te_Tphn_epsm_220_vs_30}(b) and (c) show that for the case of 30 fs pulse with $I_0 = 161$ GW/cm$^2$, the effective $T_e$ is around 200 K lower and $T_{ph}$ is a few K lower than the results of the ANTH$\mathrsfso{E}$M. This occurs because comparing with the ANTH$\mathrsfso{E}$M, the instantaneous thermalization assumption in the TTM causes the change of $\varepsilon'$ to be slightly larger (see Figs.~\ref{fig:fe_Te_Tphn_epsm_220_vs_30}(d)), resulting in a somewhat smaller absorption. This indicates that the instantaneous thermalization assumption in TTM becomes less valid for shorter pulses.

To further understand the role of the non-thermal electron distribution on the nonlinear optical response of ITO, we compare the ANTH$\mathrsfso{E}$M with the eTTM (see details in Appendix~\ref{app:thm_eps}). In contrast to the TTM, the eTTM accounts for the finite $e$-$e$ relaxation time and thus allows the electron subsystem to be non-thermal. In this model, the electron subsystem is described by an {\em instantaneous} temperature that represents only the thermal part of the electron distribution. {\XYZ This approach is essential when the $e$-$e$ thermalization time is comparable with the $e$-$ph$ relaxation time, see Refs.~\cite{non_eq_model_Lagendijk,hot_electrons_graphite_Ishida}}.  Fig.~\ref{fig:Te_epsm_dym_30_vs_ettm} shows that the results of the eTTM are in remarkable agreement with the ANTH$\mathrsfso{E}$M except that the rise time of the instantaneous $T_e$ is controlled by the $e$-$e$ relaxation time instead of the pulse duration and thus is longer than that of the extracted $T_e$ (see discussion in Appendix~\ref{app:thm_eps}). Therefore, the response time of $\varepsilon'$ is controlled by the pulse duration due to the account for the non-thermal distribution in the permittivity calculation. Moreover, if the non-thermal contribution is neglected when calculating the permittivity, the response time of $\varepsilon'$ will be controlled by the $e$-$e$ relaxation time, see Fig.~\ref{fig:Te_epsm_dym_30_vs_ettm}(b). As a result, the absorption (Fig.~\ref{fig:Te_epsm_dym_30_vs_ettm}(c)) and thus the maximum value of the instantaneous $T_e$ (Fig.~\ref{fig:Te_epsm_dym_30_vs_ettm}(a)) are much higher compared with the results accounting for the non-thermal distribution in the permittivity calculation. This implies that the instantaneous rise time of $\varepsilon’$ is due to the change of the population, as claimed in~\cite{Sapienza_double_slit_2022}. Therefore, the account for the non-thermal electron distribution is an absolute requirement for the accurate description of the nonlinear response of ITO on the few fs timescale. 

From the comparison above one can also see that comparing with the TTM, although the eTTM is more accurate, the result of the eTTM is only slightly better. {\XYZ This is quite different from noble metals~\cite{non_eq_model_Lagendijk} and graphite~\cite{hot_electrons_graphite_Ishida}.} The main reason for that is the short $e$-$e$ relaxation time in ITO. This justifies the claim for the superiority of the TTM over the eTTM in~\cite{de_Leon_Mexicans_2022_1}.

\section{Discussion}\label{sec:discussion}
{\XYZ Our model is based on a simpler model that we developed for (parabolic band) metals \cite{Dubi-Sivan, Dubi-Sivan-Faraday} and semiconductors \cite{Sarkar-Un-Sivan-Dubi-NESS-SC}. In that simpler model, due to weak illumination intensity, we didn't incorporate a self-consistent solution of field, distribution, and permittivity. Our model can also be generalized to two-dimensional materials (for example, graphene, the surface of three-dimensional topological insulators~\cite{massicotte2021hot,hot_electrons_graphite_Ishida,hasan2011three}) with proper modifications of the $e$-$ph$ and $e$-$e$ interaction terms.} This work is a starting point for the study of the nonlinear optical response of ITO at shorter wavelengths (e.g.,~\cite{Guo_ITO_nanorod_natphoton}), nonlinear effects requiring very high nonlinearity such as bistability, and to the study of the response at higher intensities and shorter pulses, and the {\XYZ formulation can also be directly implemented to} study the nonlinear optical response of other transparent conductive oxides such as TiN, ZrN,~\cite{Shalaev_Schaller_adom} etc. Our present work, along with the previous results from Ref.~\cite{Un-Sarkar-Sivan-LEDD-I}, provide a thorough understanding of the electron dynamics and the associated nonlinear response. This can further aid in understanding the benefits and limitations of transparent conductive oxides for a list of applications~\cite{Guru_Boltasseva_alternatives,Li_review_ENZ_analysis} such as pulse shaping and optical switching, efficient frequency conversion, Terahertz emission, etc. ITO being a CMOS-compatible material, all these important applications are directly relevant to our findings. Moreover, the improved understanding of the heat dynamics in ITO systems is crucial to optical communication systems concerning the development of modern data centers. %heat management in optical communication systems is a major concern for the development of modern data centers. As such, .

The non-thermal (and the simpler thermal) permittivity model described above are frequently replaced by a simpler approach - based on assigning a local-like $\chi^{(3)}$ (or $n_2$) value for the ITO nonlinearity, see, e.g.,~\cite{Boyd_NLO_ENZ_ITO,Boyd_NLO_ENZ_analysis}, or later in~\cite{Xian_group_ITO_2019,Exeter_Nat_Comm_2021,de_Leon_Mexicans_2022_1}). However, it is well-known that for absorptive materials such as metals (and hence, LEDDs), the response is strongly non-local in time, such that the fitted values strongly depend on the pulse duration, see e.g.,~\cite{Biancalana_NJP_2012,Boyd-metal-nlty}). As a particular example, Fig.~\ref{fig:fe_Te_Tphn_epsm_220_vs_30} also shows that if the pump peak intensity is reduced to 22 GW/cm$^2$ (by a factor $\sim 8$) but the pulse duration remains 30 fs, the changes of the extracted $T_e$ and $\varepsilon'$ are only reduced by 50\%. The inclusion of higher-order nonlinear coefficients cannot overcome this difficulty. Moreover, the local nonlinear susceptibility values cannot capture the asymmetric temporal response (namely, the smeared nonlinearity turn-off, see Fig.~\ref{fig:fe_Te_Tphn_epsm_220_vs_30}(d)-(e)). This prevents one from being able to directly connect the permittivity to the local field intensity, or from being able to characterize the ultrafast nonlinear response of ITO by Kerr-like nonlinearity with an intensity-dependent refractive index, e.g., in~\cite{Boyd_NLO_ENZ_ITO,Zhou_freq_trans_ENZ_2020,Exeter_Nat_Comm_2021}.

\section{Conclusions}\label{sec:conclusions}
We have presented a fully non-thermal non-phenomenological microscopic model for the electron permittivity of LEDD materials. Our work resolved several arguments and explained several experimental observations reported in LEDD material literature that was so far not understood, namely, the importance of momentum conservation in $e$-$ph$ interactions, the scaling of the collision rate with $T_e$, and the negligible energy transfer from non-thermal electrons to phonons. Our model also established the instantaneous nature of the pump-induced permittivity changes, and revealed the possibility of the chemical potential to transiently decrease dramatically, even to negative values. {\XYZ From our analysis we conclude} that the nature of the nonlinear optical response of ITO (and LEDD materials) is not polynomial nor saturable, but rather a complex temporally-local non-equilibrium response that can be approximated by thermal models only for sufficiently long pulses.

%{\bf IW -does this paragraph belong to the discussion?} 

\appendix
\section{The adiabatic approximation}\label{app:adiab_approx}
To solve Maxwell equations for systems whose material properties are time-dependent, we first derive the polarization density by solving the density matrix equations in the time domain and take the trace of the product of the transition dipole moment with the density matrix (see, e.g.,~\cite{Boyd-book}). The electric displacement $\bm{\mathscr{D}}(t)$ is then related to the local electric field $\bm{\mathscr{E}}(t)$ by~\cite{orfanidis2002electromagnetic,landau2013electrodynamics}
\begin{align}\label{eq:D_vs_E}
\bm{\mathscr{D}}(t) = \varepsilon_0\varepsilon_\infty \bm{\mathscr{E}}(t) + \int_{-\infty}^t R(t,t') \bm{\mathscr{E}}(t') dt',
\end{align}
where $R(t,t')$ is the memory function and is related to the electron distribution by
%\begin{widetext}
\begin{multline}\label{eq:R_t}
R(t,t') = \lim_{{\bf q} \rightarrow 0}\dfrac{2e^2}{i\hbar q^2} \int \dfrac{d^3 k}{(2\pi)^3} (f_{{\bf k} + {\bf q}}(t') - f_{{\bf k}}(t')) \\
e^{-[(\eta_{{\bf k} + {\bf q}}(f(t')) + \eta_{{\bf k}}(f(t')))/2 + i (\e_{{\bf k} + {\bf q}} - \e_{{\bf k}})/\hbar](t-t')} .
\end{multline}
%\end{widetext}
If the damping rate $\eta_{\bf k}$ is much faster than the rate of change of the electron distribution, only the electron dynamics at $t'$ nearby $t$ contributes to the integral in Eq.~\eqref{eq:D_vs_E}. Since the change rate of the electron distribution is caused by absorption via the electron excitation term, which is incoherent in nature (i.e., it depends on $|\mathscr{E}|^2$), the condition above is satisfied when $\left|\left(df/dt\right)_\text{exc}\right| < \eta/\hbar$. This requires the local field to be smaller than $2.7 \times 10^9$ V/m~\footnote{For ITO, the damping rate is around 0.1 fs$^{-1}$, see Fig. 1 in~\cite{Un-Sarkar-Sivan-LEDD-I}. From Eqs.~\eqref{eq:dfEdt_exc}-\eqref{eq:exc_eng_conserv}, it follows that the condition $\left|\left(df/dt\right)_{exc}\right| < \eta$ is therefore satisfied when the local field is smaller than $\sim 2.7 \times 10^9$ V/m, corresponding to an incident intensity of $\sim 200$ GW/cm$^2$ for the example analyzed in this work. Since the absorption decreases as $T_e$ increases, a much higher incident intensity than $\sim 200$ GW/cm$^2$ is required to violate $\left|\left(df/dt\right)_\text{exc}\right| < \eta$ at higher $T_e$.}. In this case, one can replace $f_{{\bf k}}(t')$ by $f_{{\bf k}}(t)$ and $\eta_{{\bf k}}(f(t'))$ by $\eta_{{\bf k}}(f(t))$, so that we can factor the electron distribution out of the integral. We further assume that the pulse duration is much longer than the periodicity of the carrier wave ($\tau_\text{pump} \gg 2\pi/\omega_\text{pump}$) and is also much longer than the damping time ($\tau_\text{pump} \gg 1/\eta$)%{\bf IW - isn't this the same as the first assumption? so avoid the repetition... - not quite. the first assumption mainly concerns the local electric field while the assumption here concerns the pump pulse duration.}
. The condition $\tau_\text{pump} \gg 2\pi/\omega_\text{pump}$ allows us to write the electric field (and the electric displacement) as a product of slowly varying envelope (${\bf E}(t')$ or ${\bf D}(t')$) and a rapidly varying phase factor with the carrier frequency, i.e., $\bm{\mathscr{E}}(t') = {\bf E}(t') e^{-i\omega_\text{pump}t'} + \text{c.c.}$ ($\bm{\mathscr{D}}(t') = {\bf D}(t') e^{-i\omega_\text{pump}t'} + \text{c.c.}$). After substituting $\bm{\mathscr{E}}(t')$ back to Eq.~\eqref{eq:D_vs_E}, the condition $\tau_\text{pump} \gg 1/\eta$ allows us to replace ${\bf E}(t')$ by ${\bf E}(t)$ so that the envelope function ${\bf E}(t)$ can be factored out of the integral. Then, Eq.~\eqref{eq:D_vs_E} becomes 
\begin{widetext}
\begin{align}\label{eq:Denv_vs_Eenv}
{\bf D}(t) = \varepsilon_0\varepsilon_\infty {\bf E}(t) + \lim_{{\bf q} \rightarrow 0} \int \dfrac{d^3k}{(2\pi)^3} (f_{{\bf k} + {\bf q}}(t) - f_{{\bf k}}(t)) {\bf E}(t)
\dfrac{2e^2}{i \hbar q^2} \int_{-\infty}^t dt' e^{-((\eta_{{\bf k} + {\bf q}}(t) + \eta_{{\bf k}}(t))/2 + i (\e_{{\bf k} + {\bf q}} - \e_{{\bf k}})/\hbar)(t - t')}
e^{i\omega_\text{pump}(t-t')}.
\end{align}
\end{widetext}
Eq.~\eqref{eq:Denv_vs_Eenv} can be reorganized into the form ${\bf D}(t) = \varepsilon_0\varepsilon(t;\omega_\text{pump}){\bf E}(t)$ where $\varepsilon(t;\omega_\text{pump})$ is given by Eq.~\eqref{eq:eps_pump_adiab}. Comparing with Eqs.~\eqref{eq:D_vs_E} and~\eqref{eq:R_t}, the approximations~\eqref{eq:Denv_vs_Eenv} and~\eqref{eq:eps_pump_adiab} state that the time-dependent permittivity changes with electron distribution adiabatically. 

When substituting the electric displacement field into the Amp\`{e}re's law, the time derivative of the electric displacement field becomes
\begin{multline*}
\dfrac{\partial \bm{\mathscr{D}}(t)}{\partial t} = \varepsilon_0\Bigg[\dfrac{\partial \varepsilon(t;\omega_\text{pump})}{\partial t} {\bf E}(t) +
\varepsilon(t;\omega_\text{pump})\dfrac{\partial {\bf E}(t)}{\partial t} \\-i\omega_\text{pump} \varepsilon(t;\omega_\text{pump}){\bf E}(t)\Bigg] e^{-i\omega_\text{pump}t} + \text{c.c.}.
\end{multline*}
Since the oscillation of the carrier wave is much {faster} than the change rate of the envelope ${\bf E}(t)$ and of $\varepsilon(t;\omega_\text{pump})$, the time derivative of the electric displacement field can be approximated by
$\dfrac{\partial \bm{\mathscr{D}}(t)}{\partial t} \approx -i\omega_\text{pump} \varepsilon_0\varepsilon(t;\omega_\text{pump}){\bf E}(t) e^{-i\omega_\text{pump}t} + \text{c.c.}$ so that the Amp\`{e}re's law becomes $\nabla\times{\bf H}({\bf r},t) = - i \omega_\text{pump} \varepsilon_0 \varepsilon(t;\omega_\text{pump}) {\bf E}({\bf r},t)$. Similarly, the Maxwell-Faraday law becomes $\nabla\times{\bf E}({\bf r},t) = - i \omega_\text{pump} \mu_0 {\bf H}({\bf r},t)$, see Eq.~\eqref{eq:Maxwell_eq_adiab}. The solution of Eqs.~\eqref{eq:Maxwell_eq_adiab} and the time-dependent permittivity~\eqref{eq:eps_pump_adiab} are then used to calculate the power absorbed density via the Ponyting theorem, see Eq.~\eqref{eq:P_abs}. When the pulse duration is comparable to the carrier wave periodicity, one should go beyond the adiabatic approximation and calculate the absorbed power density using  Poynting's theorem in the time domain.

\section{Extended two-temperature model (eTTM) and two-temperature model (TTM)}~\label{app:eTTMnTTM}
Following the approach in~\cite{Dubi-Sivan,Dubi-Sivan-Faraday}, the dynamics of the electron and phonon subsystems can be macroscopically described by deriving the extended two temperature model (eTTM) by coarse-graining the Boltzmann equation~\eqref{eq:f_neq_dynamics}. In this approach, we split the Boltzmann equation~\eqref{eq:f_neq_dynamics} for the electron distribution into a thermal and a non-thermal part, $f(\e) = f^T(\e,\mu(T_e),T_e) + f^{NT}(\e)$, and add the energy balancing equation for the phonon subsystem. We then multiply the Boltzmann equation by the product of the electron energy, $\e$, and the density of electron states, $\rho_e(\e)$, and integrate over the electron energy. This results in a pair of equations describing the dynamics of the non-thermal electron distribution $f^{NT}$ and its integral version (i.e., an equation for the non-thermal electron energy $\U^{NT} = \displaystyle\int \e \rho_e(\e)f^{NT}d\e$), 
\begin{subequations}\label{eq:eTTM_NT}
\begin{align}
\dfrac{\partial f^{NT}}{\partial t}&=\left(\dfrac{\partial f}{\partial t}\right)_{exc} + \left(\dfrac{\partial f^{NT}}{\partial t}\right)_{e\text{-}e} + \left(\dfrac{\partial f^{NT}}{\partial t}\right)_{e\text{-}ph}, \label{eq:eTTM_fNT}\\
\dfrac{d \U^{NT}}{d t}&=p_\text{abs}(t) + \left(\dfrac{d \U^{NT}}{d t}\right)_{e\text{-}e} + \left(\dfrac{d \U^{NT}}{d t}\right)_{e\text{-}ph}, \label{eq:eTTM_UNT}
\end{align}
\end{subequations}
as well as equations for the thermal electron energy,
\begin{align}
C_e(T_e)\dfrac{dT_e}{dt}&=\left(\dfrac{d\U^{T}}{dt}\right)_{e\text{-}ph}-\left(\dfrac{d\U^{NT}}{dt}\right)_{e\text{-}e},
\label{eq:eTTM_Te}
\end{align}
and phonon energy
\begin{align}
C_{ph}\dfrac{dT_{ph}}{dt}&=-\left(\dfrac{d\U^{T}}{dt}\right)_{e\text{-}ph} -\left(\dfrac{d\U^{NT}}{dt}\right)_{e\text{-}ph}. \label{eq:eTTM_Tph} 
\end{align}
$C_e$ is the electron heat capacity which can be determined by (see details in~\cite{Un-Sarkar-Sivan-LEDD-I})
\begin{align}
C_e(T_e) = \int \e \rho_e(\e)\dfrac{\partial f^T(\e,\mu(T_e),T_e)}{\partial T_e} d\e,
\end{align}
where $\mu(T_e)$ is the $T_e$-dependent chemical potential~\cite{Un-Sarkar-Sivan-LEDD-I} which ensures the number conservation of the electron subsystem, 
\begin{align}\label{eq:num_conserv_mu}
\int\rho_e(\e)f^T(\e,\mu(T_e),T_e)d\e = \int\rho_e(\e)f^T(\e,\e_F,0\ \textrm{K})d\e.
\end{align}
The phonon heat capacity is taken to be $C_{ph} = 2.54 \times 10^6$ J m$^{-3}$ K$^{-1}$~\cite{ITO_Cph}. 

$T_e$ in Eq.~\eqref{eq:eTTM_Te} is referred to as the {\XYZ {\em instantaneous} electron temperature}; it differs from the temperature extracted from the solution of BE (see Fig.~\ref{fig:Te_epsm_dym_30_vs_ettm} in Appendix~\ref{app:thm_eps}) because the {\em instantaneous} $T_e$ has a longer rise time than the {\em extracted} $T_e$. $({\partial f}/{dt})_\text{exc}$ is the electron-photon excitation term given by Eq.~\eqref{eq:dfEdt_exc}, while $(\partial f^{NT}/\partial t)_{e\text{-}e}$ and $(\partial f^{NT}/\partial t)_{e\text{-}ph}$ are, respectively, the rates of non-thermal electron relaxation via the $e$-$e$ and $e$-$ph$ interactions. $p_\text{abs}(t)$ is the absorbed power density given by Eq.~\eqref{eq:P_abs}, $(d \U^{NT}/d t)_{e\text{-}e}$ is the non-thermal energy relaxation due to the $e$-$e$ interaction; $(d \U^{NT}/d t)_{e\text{-}ph}$ and $\left({d\U^{T}}/{dt}\right)_{e\text{-}ph}$ are, respectively, the energy transfer from the non-thermal and thermal electrons to phonons, and $C_e$ and $C_{ph}$ are, respectively, the heat capacity of the electron and phonon subsystems. In Eq.~\eqref{eq:eTTM_fNT}, we calculate the non-thermal part of the electron distribution explicitly since the non-thermal energy is not enough for calculating the permittivity, see Appendix~\ref{app:thm_eps}.

%\onecolumngrid
%\begin{center}
\begin{figure*}[ht]
\centering
\includegraphics[width=0.9\textwidth]{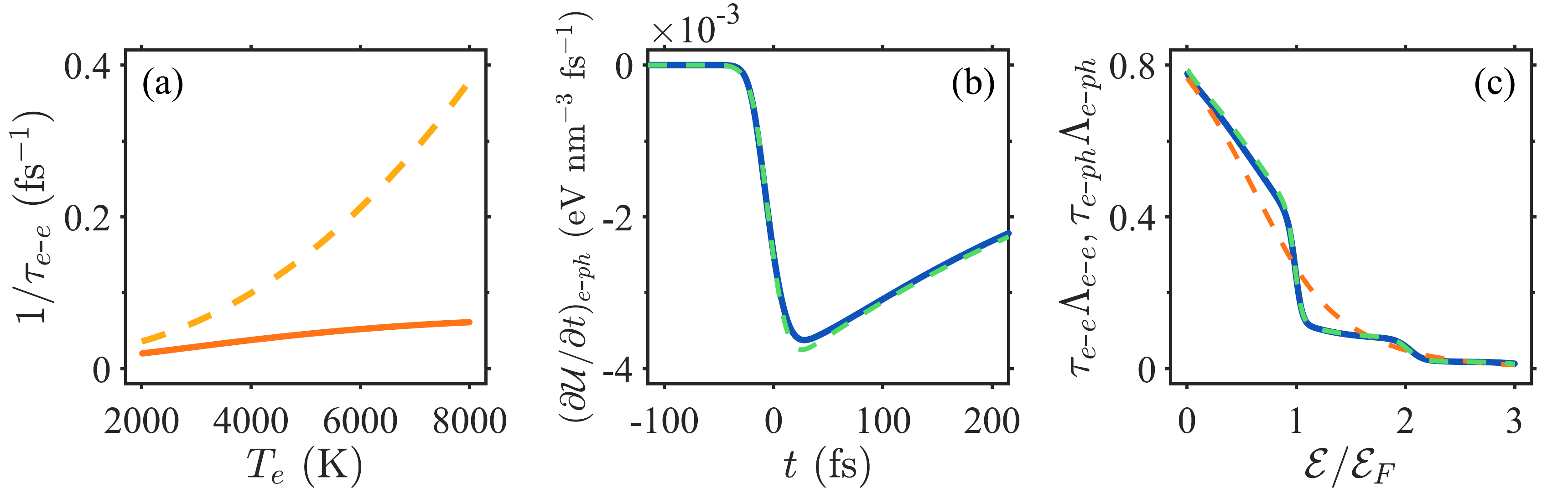}
\caption{(Color online) (a) The $T_e$-dependence of the $e$-$e$ relaxation rate $\tau_{e\text{-}e}^{-1}$. The orange solid line represents the result calculated using Eq.~(17) in~\cite{Un-Sarkar-Sivan-LEDD-I} at $\e = \mu(T_e) + \hbar\omega_\text{pump}/2$. The orange dashed line represents the result calculated using the Gurzhi formula~\eqref{eq:ee_rel_Gurzhi}. (b) The rate of total energy transfer from the electron to the phonon subsystem Eq.~\eqref{eq:dUdt_eph} (blue solid line). The green dashed line is the contribution from the thermal part of the electron distribution~\eqref{eq:dUTdt_eph}. The results correspond to case of $I_0 = 161$ GW/cm$^2$ and $\tau_\text{pump} = 30$ fs. (c) The electron energy dependence of $\Lambda_{e\text{-}ph}\tau_{e\text{-}ph}$ (see Eq.~\eqref{eq:ephn_RTA}) (green dashed line) and $\Lambda_{e\text{-}e}\tau_{e\text{-}e}$ %{\bf IW - should be products?} 
(orange dashed line) for the same case as (b) at $t = 0$ fs. The blue solid line is the non-equilibrium electron distribution {\XYZ in Eq.~\eqref{eq:ephn_RTA} at $t = 0$ fs}. %before relaxation {\bf IW - so at what $t$? is that the $f$ appearing in B9?}. 
}
\label{fig:tau_ee_ephn}
\end{figure*}
%\end{center}
%\twocolumngrid

Unlike the rigorous approach of Eq.~(\ref{eq:f_neq_dynamics}), to derive the eTTM, the $e$-$e$ relaxation term is simplified using the relaxation time approximation, namely, 
\begin{align}
\left(\dfrac{\partial f^{NT}}{\partial t}\right)_{e\text{-}e} = -\dfrac{f^{NT}}{\tau_{e\text{-}e}},
\end{align}
where $\tau_{e\text{-}e}$ is the $e$-$e$ relaxation time. In many previous studies~\cite{non_eq_model_Carpene,Boyd_NLO_ENZ_ITO,Boyd_Nat_Phot_2018,de_Leon_Mexicans_2022_1,Ellenbogen-Minerbi-ITO}, $\tau_{e\text{-}e}$ was estimated using the Gurzhi formula~\cite{Gurzhi_ee_rel,Smith_Ehrenreich_PRB_1982,Del_Fatti_ultrafast_NLTY_JPCB_2001},
\begin{align}\label{eq:ee_rel_Gurzhi}
\dfrac{1}{\tau_{e\text{-}e}(T_e)} = \dfrac{\omega_\text{pump}^2}{4\pi^2\omega_p(T_e)}\left[1 + \left(\dfrac{2\pi k_B T_e}{\hbar\omega_\text{pump}}\right)^2\right],
\end{align}
where $\omega_p(T_e)$ is the plasma frequency given by Eq.~\eqref{eq:omgp_sq}~\cite{Guo_ITO_AM_2017,Guo_ITO_nanorod_natphoton,Boyd_NLO_ENZ_ITO,Boyd_Nat_Phot_2018,Kinsey_ENZ_OptMatExp,Xian_group_ITO_2020,Exeter_Nat_Comm_2021,Exeter_Nat_Comm_2021,Sapienza_2022,de_Leon_Mexicans_2022_1}. This formula is identical to the Fermi-liquid relaxation time with $\e = \e_F + \hbar\omega/2$; it is a good approximation for high electron density Drude materials such as noble metals. For LEDD materials with a non-parabolic conduction band such as ITO, the Gurzhi formula~\eqref{eq:ee_rel_Gurzhi} suggests that $\tau_{e\text{-}e}^{-1}$ exhibits a super-quadratic growth with $T_e$ (since the plasma frequency Eq.~\eqref{eq:omgp_sq} decreases with $T_e$~\cite{Guo_ITO_nanorod_natphoton,Guo_ITO_nanorods_NC_2016,Boyd_Nat_Phot_2018,Ellenbogen-Minerbi-ITO,de_Leon_Mexicans_2022_1}). In contrast, in the ANTH$\mathrsfso{E}$M, we calculate the $e$-$e$ relaxation time from the functional derivative of the $e$-$e$ collision term. In particular, $\tau_{e\text{-}e}$ is shown~\cite{Un-Sarkar-Sivan-LEDD-I} to be well described by the Fermi-liquid theory with a correction factor, i.e., $\tau^{-1}_{e\text{-}e,{\bf k}} \sim (1 + 2 C \mu(T_e))^3\left[(\pi k_B T_e)^2 + (\e_{{\bf k}} - \mu(T_e))^2\right]$. Due to the decrease of the chemical potential with $T_e$~\cite{Un-Sarkar-Sivan-LEDD-I} (see also Fig.~\ref{fig:Te_Tph_dym}(b)), it increases only nearly linearly with $T_e$ (hence, much slower than that predicted by the Gurzhi formula). This finding justifies the phenomenological adjustment to the $T_e$-dependence of $\tau_{e\text{-}e}^{-1}$ in~\cite{de_Leon_Mexicans_2022_1}. 

$(\partial f^{NT}/\partial t)_{e\text{-}ph}$ and $(d \U^{NT}/d t)_{e\text{-}ph}$ were also treated using the relaxation time approximation in previous studies~\cite{Boyd_NLO_ENZ_ITO,Boyd_Nat_Phot_2018,Ellenbogen-Minerbi-ITO}, i.e., $(\partial f^{NT}/\partial t)_{e\text{-}ph} = - f^{NT}/\tau_{e\text{-}ph}$ and $(d\U^{NT}/d t)_{e\text{-}ph} = -\U^{NT}/\tau_{e\text{-}ph}$, where $\tau_{e\text{-}ph}$ is the $e$-$ph$ relaxation time and was estimated using (e.g.,~\cite{Boyd_Nat_Phot_2018})
\begin{align}\label{eq:eph_rel_Boyd}
\dfrac{1}{\tau_{e\text{-}ph}(T_{ph})} = \Gamma_0 \left[\dfrac{2}{5} + \dfrac{4T_{ph}^5}{T_D^5} \int_0^{T_D/T_{ph}}\dfrac{z^4}{e^z-1}dz\right],
\end{align} 
where $\Gamma_0 = 0.5342$ fs$^{-1}$, and $T_D = 900$ K is the Debye temperature. In the ANTH$\mathrsfso{E}$M, $\tau_{e\text{-}ph}$ is evaluated from the functional derivative of the $e$-$ph$ collision term accounting for the momentum conservation, and is shown to be proportional to the phonon temperature and to be weakly dependent on $T_e$. The value of $\tau_{e\text{-}ph}$ evaluated using the explicit form given in~\cite{Un-Sarkar-Sivan-LEDD-I} is similar to that obtained from Eq.~\eqref{eq:eph_rel_Boyd} and is shorter than 10 fs, i.e., much shorter than the $e$-$e$ relaxation time. In this case, most of the non-thermal energy directly relaxes to the phonon subsystem after the generation of the non-thermal electron distribution, such that the electron subsystem will not heat up at all. This contradicts with the experimental observations which typically conclude that the $e$-$ph$ relaxation is much slower than the $e$-$e$ relaxation (e.g.,~\cite{de_Leon_Mexicans_2022_1}). This disagreement can be reconciled by comparing the total energy transfer rate from electrons to phonons (denoted as $(d\U/dt)_{e\text{-}ph}$) with its contribution from the thermal electrons (i.e., $(d\U^T/dt)_{e\text{-}ph}$). These quantities can be calculated from the $e$-$ph$ collision term in Eq.~\eqref{eq:f_neq_dynamics} using the (total) electron distribution and a Fermi-Dirac distribution, namely,
\begin{align}\label{eq:dUdt_eph}
\left(\dfrac{d\U}{dt}\right)_{e\text{-}ph} = \displaystyle\int \e\rho_e(\e) \left(\dfrac{\partial f}{\partial t}\right)_{e\text{-}ph}d\e
\end{align}
and 
\begin{align}\label{eq:dUTdt_eph}
\left(\dfrac{d\U^{T}}{dt}\right)_{e\text{-}ph} = \displaystyle\int \e\rho_e(\e) \left(\dfrac{\partial f^T(\e,\mu(T_e),T_e)}{\partial t}\right)_{e\text{-}ph}d\e.
\end{align} $(d\U^{NT}/dt)_{e\text{-}ph}$ is then the difference between $(d\U/dt)_{e\text{-}ph}$ and $(d\U^{T}/dt)_{e\text{-}ph}$, i.e., $$\left(\dfrac{d\U^{NT}}{dt}\right)_{e\text{-}ph} = \left(\dfrac{d\U}{dt}\right)_{e\text{-}ph} - \left(\dfrac{d\U^T}{dt}\right)_{e\text{-}ph}.$$ 
Our model shows that the total $e$-$ph$ energy transfer rate is dominated by the thermal part of the electrons, i.e., $(d\U/dt)_{e\text{-}ph}\approx(d\U^{T}/dt)_{e\text{-}ph}$, see Fig.~\ref{fig:tau_ee_ephn}(b). This means that the contribution from the non-thermal part of the electrons $(d\U^{NT}/dt)_{e\text{-}ph}$ is, in fact, negligible (see also~\cite{Dubi-Sivan,Dubi-Sivan-Faraday,de_Leon_Mexicans_2022_1}), confirming the ad hoc neglect of $(d\U^{NT}/dt)_{e\text{-}ph}$ done in~\cite{de_Leon_Mexicans_2022_1}.

To have a deeper understanding of this, we cast the $e$-$ph$ collision term into an RTA form that enforces number conservation (by adding the Lorentz term, see~\cite{Dubi-Sivan}), namely,  
\begin{align}\label{eq:ephn_RTA}
(\partial f/\partial t)_{e\text{-}ph} = -f/\tau_{e\text{-}ph} + \Lambda_{e\text{-}ph}.
\end{align} 
Thus, $\tau_{e\text{-}ph}\Lambda_{e\text{-}ph}$ can be interpreted as the target distribution which the non-equilibrium electron distribution is going to relax to via the $e$-$ph$ interaction if the excitation is stopped and the $e$-$e$ interaction is turned off. Fig.~\ref{fig:tau_ee_ephn}(c) shows that the target distribution {\XYZ has a non-thermal shoulder structure and differs from the distribution before relaxation only slightly, thus is distinct from any thermal distribution.} %{\bf not to me... maybe trace a thermal distribution? or maybe this is not so important?}
In contrast, if we perform the same analysis for the $e$-$e$ collision term, we find that the target distribution is similar to a thermal distribution with some high temperature, see Fig.~\ref{fig:tau_ee_ephn}(c). Moreover, the $e$-$e$ Lorentz term is two orders of magnitude larger than the $e$-$ph$ Lorentz term. This explains why the $e$-$e$ relaxation is much faster than the $e$-$ph$ relaxation ($\big|\left(\partial f/\partial t\right)_{e\text{-}e}\big| \gg \big|\left(\partial f/\partial t\right)_{e\text{-}ph}\big|$, see Fig.~2 in~\cite{Un-Sarkar-Sivan-LEDD-I}) although $\tau_{e\text{-}e}^{-1} < \tau_{e\text{-}ph}^{-1}$. The comparison above further indicates that the $e$-$ph$ interaction is mainly responsible for transferring the thermal energy of the electrons to phonon, instead of thermalizing the non-equilibrium electron distribution. Therefore, the $e$-$ph$ interaction violates the RTA assumption, so that it is improper to apply the relaxation time approximation for the $e$-$ph$ relaxation of the non-thermal electron energy. This conclusion is also valid for noble metals. 

The energy transfer rate from the thermal electrons to the phonons $\left({d\U^{T}}/{dt}\right)_{e\text{-}ph}$ (Eq.~\eqref{eq:dUTdt_eph}) has been shown in~\cite{Un-Sarkar-Sivan-LEDD-I} to be proportional to the temperature difference between electrons and phonons, namely, $\left({d\U^{T}}/{dt}\right)_{e\text{-}ph} = -G_{e\text{-}ph}(T_e-T_{ph})$, where $G_{e\text{-}ph}$ is the electron-phonon energy coupling coefficient. The electron-phonon energy coupling coefficient was previously evaluated using the formulation derived for noble metals (e.g., in~\cite{Boyd_Nat_Phot_2018,de_Leon_Mexicans_2022_1,Ellenbogen-Minerbi-ITO}. However, we showed in~\cite{Un-Sarkar-Sivan-LEDD-I} that this approach overestimates the correct value by more than an order of magnitude due to the failure to account for momentum conservation in the $e$-$ph$ interaction. This explains the need to reduce the $e$-$ph$ coupling coefficient by a factor of 8 in~\cite{de_Leon_Mexicans_2022_1} to fit the experimental data.

%{\bf IW - discuss this before the discusison of G?} 

Finally, if the $e$-$e$ relaxation time is approximated to be zero, i.e., the electron subsystem is assumed to become thermalized instantaneously, Eq.~\eqref{eq:eTTM_NT} reduces to the two-temperature model (TTM)~\cite{Two_temp_model},
\begin{subequations}\label{eq:TTM}
\begin{align}
C_e(T_e)\dfrac{dT_e}{dt} &= - G_{e\text{-}ph}(T_e-T_{ph}) + P_\text{abs}(t), \label{eq:TTM_Te} \\
C_{ph}\dfrac{dT_{ph}}{dt} &= G_{e\text{-}ph}(T_e-T_{ph}). \label{eq:TTM_Tph}
\end{align}
\end{subequations}
Here, $T_e$ is referred to as the {\em effective} electron temperature; it differs from %In general, the {\em effective} $T_e$ is different from 
the {\em extracted} and {\em instantaneous} $T_e$  since thermalization is assumed to be instantaneous in Eq.~(\ref{eq:TTM}).

As shown in this work and others~\cite{Un-Sarkar-Sivan-LEDD-I,de_Leon_Mexicans_2022_1}, the TTM can be quite accurate for pulses having duration longer than the $e$-$e$ relaxation time. 

\section{Limitation of using thermal distribution to evaluate the permittivity }\label{app:thm_eps}
%\onecolumngrid
%\begin{center}
\begin{figure*}[t]
\centering
\includegraphics[width=0.9\textwidth]{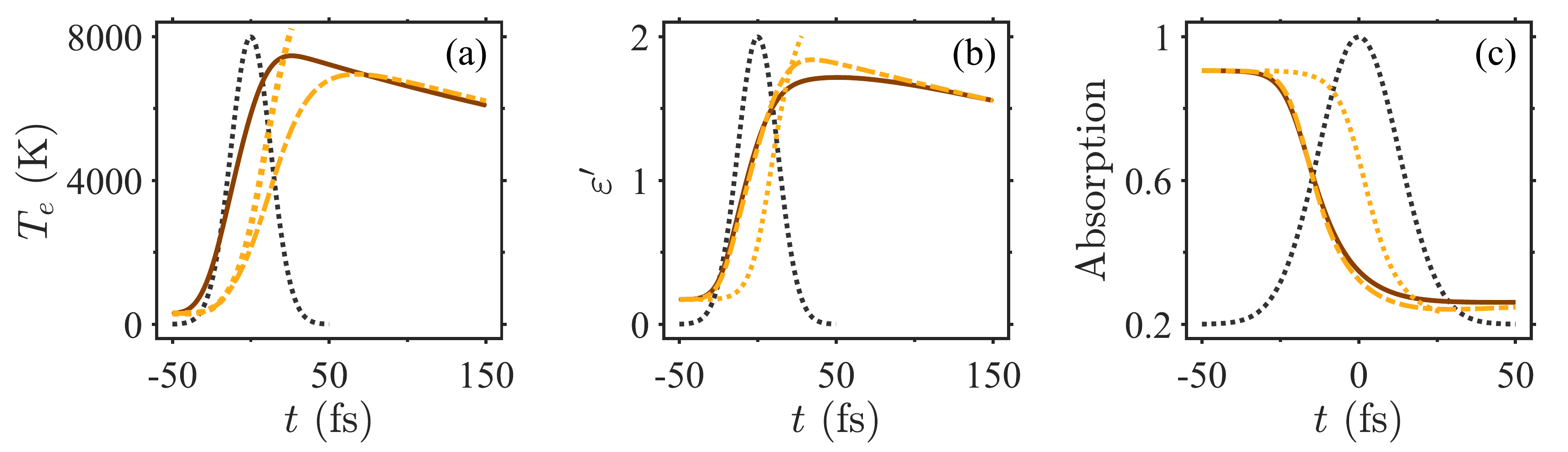}
\caption{(Color online) (a) and (b) the same as Fig.~\ref{fig:fe_Te_Tphn_epsm_220_vs_30}(b) and (d) for the 30 fs pulse with $I_0 = 161$ GW/cm$^2$. (c) The absorption as a function of time for the same case as (a) and (b). The dash-dotted and dotted lines represent the results obtained from eTTM with non-thermal and thermal permittivity model, respectively. The black dotted lines represent the temporal intensity profile of the incident pump pulse.}
\label{fig:Te_epsm_dym_30_vs_ettm}
\end{figure*}
%\end{center}
%\twocolumngrid
In this section, we use the example of the 30 fs pump pulse with $I_0 = 161$ GW/cm$^2$ studied in Section~\ref{subsec:short_pulse} to demonstrate the importance of accounting for the non-thermal part of the electron distribution in the permittivity calculation when self-consistently solving the eTTM (Eqs.~\eqref{eq:eTTM_NT}-\eqref{eq:eTTM_Tph}) coupled with Maxwell's equations~\eqref{eq:Maxwell_eq_adiab}. To do that, we evaluate the permittivity using two different methods in the self-consistent calculation. In the first case, the ITO permittivity (Eq.~\eqref{eq:eps_pump_adiab}) is evaluated using the electron distribution $f(\e) = f^T(\e,\mu(T_e),T_e) + f^{NT}(\e)$. In the second one, we neglect $f^{NT}(\e)$ and evaluate the ITO permittivity using only the thermal distribution $f(\e,\mu(T_e),T_e)$. 
For the first case, the instantaneous $T_e$ dynamics is qualitatively similar to that of the extracted $T_e$ except that at the early stages, the rise of the instantaneous $T_e$ is controlled by $e$-$e$ relaxation time, which is slower than the pulse duration, see Fig.~\ref{fig:Te_epsm_dym_30_vs_ettm}(a). In contrast, the rise time of $\varepsilon'$ (and thus the decrease of absorption) are controlled by the pulse duration, showing remarkable agreement with the results obtained from our ANTH$\mathrsfso{E}$M (see Fig.~\ref{fig:Te_epsm_dym_30_vs_ettm}(b) and (c)). 

However, if one neglects the non-thermal part of the electron distribution and simply evaluates the ITO permittivity using a thermal distribution with %{\bf IW - instantaneous/extracted?} 
{\XYZ instantaneous} $T_e$, the rise time of the permittivity (and the decrease of absorption) will be the same as {\XYZ instantaneous} $T_e$ thus will be controlled by the $e$-$e$ relaxation time instead of the pulse duration, see the dotted lines in Fig.~\ref{fig:Te_epsm_dym_30_vs_ettm}. This causes the maximum value of the instantaneous $T_e$ to be $> 10^4$ K, much higher than the result of the case accounting for the non-thermal distribution in the permittivity calculation. This comparison clearly shows that the non-thermal electron distribution is essential to properly capture the nonlinear response of ITO.

The need to account for the non-thermal distribution also occurred when calculating the change of the contribution of interband transitions to the permittivity of noble metals under ultrafast laser pulse excitation. In that case, the non-thermal electron distribution was assumed to have the same profile as the photon excitation term, i.e., $f^{NT}(\e,t) = A(t) \big[f^T(\e-\hbar\omega_\text{pump}) \left(1 - f^T(\e)\right) - f^T(\e)\left(1 - f^T(\e + \hbar\omega_\text{pump})\right)\big]$, where $A(t)$ is determined from the non-thermal energy $\U^{NT}(t)$~\cite{non_eq_model_Ippen,non_eq_model_Carpene,valle_transient_nlty_response_2015}. This is a decent approximation for noble metals because the chemical potential is nearly $T_e$-independent. Unfortunately, this approximation is not valid for ITO since the chemical potential decreases significantly with $T_e$, thus, requiring the use of the full non-thermal model (ANTH$\mathrsfso{E}$M).

\section{Probe pulse dynamics for the shorter pump pulses}\label{app:probe_dyn_short_pulse}
\begin{figure}[b]
\centering
\includegraphics[width=0.7\columnwidth]{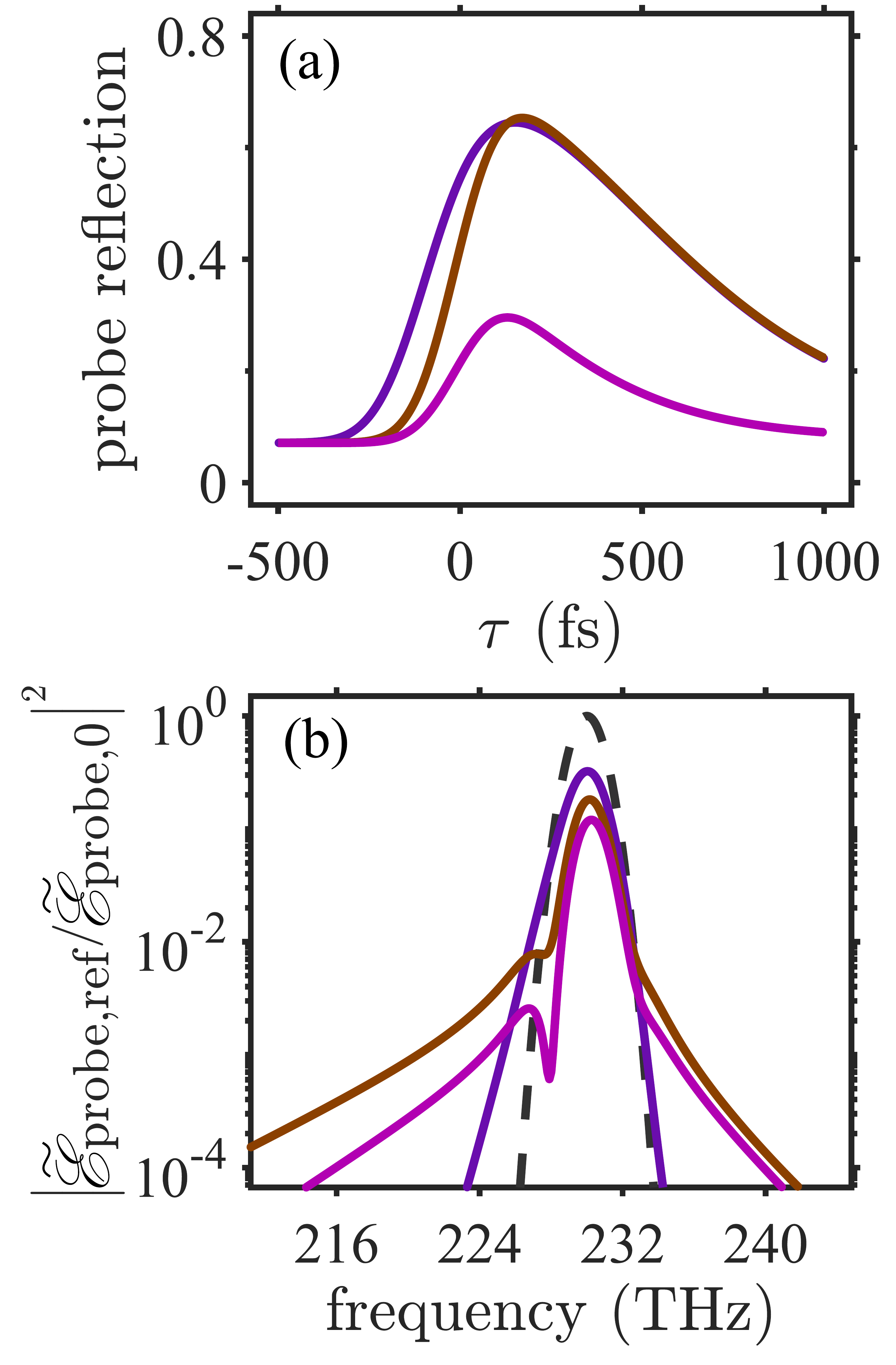}
\caption{(Color online) The same as Fig.~\ref{fig:probe_refl_decay}(a) and (c) but for cases of shorter pump pulses studied in Fig.~\ref{fig:fe_Te_Tphn_epsm_220_vs_30} (the same color is used). }\label{fig:probe_refl_decay_widen_v2_220_vs_30}
\end{figure}
{\XYZ In this Appendix, we show the comparison of the reflection and spectrum of the reflected probe pulse (with duration of 220 fs) among the three cases studied in Section~\ref{subsec:short_pulse}. Fig.~\ref{fig:probe_refl_decay_widen_v2_220_vs_30}(a) shows that decay rate of the probe pulse reflection for the case of 30 fs pump pulse with $I_0 = 161$ GW/cm$^2$ is (almost) the same as the case of 220 fs pump pulse with $I_0 = 22$ GW/cm$^2$, and is slower than that the case of 30 fs pump pulse with $I_0 = 22$ GW/cm$^2$. This indicates that the decay rate of the probe pulse reflection is slower for stronger pump pulse energy but is weakly sensitive to the pump pulse duration. In contrast, the spectrum of the reflected probe pulse is wider for shorter pump pulse duration, see Fig.~\ref{fig:probe_refl_decay_widen_v2_220_vs_30}(b). }

%shows that the spectrum of the reflected probe pulse for the cases of 30 fs is much wider than the case of 220 fs pump pulse

%for the case of 30 fs pump pulse with $I_0 = 161$ GW/cm$^2$ the reflection of the probe pulse has a faster increase rate but has (almost) the same decay rate with the time-delay comparing with the case of 220 fs pump pulse with $I_0 = 22$ GW/cm$^2$. This indicates that the increase rate of the probe pulse reflection is mainly determined by the pump pulse duration while the decay rate is mainly determined by the pump pulse energy. }

\input{Rev_dc_main.bbl}
%\bibliography{my_bib}
\end{document}

%% file: Rev_dc_main.bbl
%apsrev4-2.bst 2019-01-14 (MD) hand-edited version of apsrev4-1.bst
%Control: key (0)
%Control: author (8) initials jnrlst
%Control: editor formatted (1) identically to author
%Control: production of article title (0) allowed
%Control: page (0) single
%Control: year (1) truncated
%Control: production of eprint (0) enabled
%